\documentclass[prl,aps,twocolumn,superscriptaddress,longbibliography,nofootinbib,floatfix]{revtex4-1}

\usepackage{amsmath,graphicx,epsfig}
\usepackage{epstopdf}
\usepackage{euscript}
\usepackage{amsfonts}
\usepackage{amssymb}
\usepackage{float}
\usepackage{tabularx}
\usepackage{bm}

\usepackage[arrows]{ezedits}
\defineEdit{DK}{\color[rgb]{.7,.9,.8}}{\color[rgb]{0,.5,.3}}

\usepackage[utf8]{inputenc}

\usepackage[toc,page]{appendix}

\def\prn#1{{\left(#1\right)}}

\def\sbrk#1{{\left[#1\right]}}
\def\abrk#1{{\langle#1\rangle}}

\def\bra#1{{\langle#1|}}

\def\pdbydt#1{{\frac{\partial #1}{\partial t}}}

\def\cg(#1,#2)(#3,#4)(#5,#6){\bra{#1,#2,#3,#4}#5,#6\rangle}

\def\threej(#1,#2)(#3,#4)(#5,#6){\begin{pmatrix}#1&#3&#5\\#2&#4&#6\end{pmatrix}}
\def\sixj(#1,#2,#3)(#4,#5,#6){\begin{Bmatrix}#1&#2&#3\\#4&#5&#6\end{Bmatrix}}
\def\ninej(#1,#2,#3)(#4,#5,#6)(#7,#8,#9){\begin{Bmatrix}#1&#2&#3\\#4&#5&#6\\#7&#8&#9\end{Bmatrix}}

\def\Msun{{\ensuremath{M_\odot}}}
\def\SupMat{Supplementary Information}

\def\mr{\mathrm}
\def\mb{\mathbf}

\newlength{\defbaselineskip}
\newcommand{\setlinespacing}[1]%
           {\setlength{\baselineskip}{#1 \defbaselineskip}}

\begin{document}

\title{Quantum sensor networks as exotic field telescopes for multi-messenger astronomy} 

\date{\today}

\author{Conner Dailey}
\affiliation{Department of Physics, University of Nevada, Reno, Nevada 89557, USA}

\author{Colin Bradley}
\affiliation{Department of Physics, University of Nevada, Reno, Nevada 89557, USA}

\author{Derek F. Jackson Kimball}
\affiliation{Department of Physics, California State University - East Bay, Hayward, California 94542, USA}

\author{Ibrahim Sulai}
\affiliation{Department of Physics and Astronomy, Bucknell University, Lewisburg, Pennsylvania 17837, USA}

\author{Szymon Pustelny}
\affiliation{Institute of Physics, Jagiellonian University, Krak\'ow, Poland}

\author{Arne Wickenbrock}
\affiliation{Helmholtz Institute Mainz, Johannes Gutenberg University, 55099 Mainz, Germany}

\author{Andrei Derevianko}
\affiliation{Department of Physics, University of Nevada, Reno, Nevada 89557, USA}

\maketitle


\textbf{Multi-messenger astronomy, the coordinated observation of different classes of signals originating from the same astrophysical event, provides a wealth of information about astrophysical processes with far-reaching implications~\cite{abbott2017multi,Alexeyev1988,Kepko2009,Aartsen2018}. So far, the focus of multi-messenger astronomy has been the search for conventional signals from known fundamental forces and standard model particles, like gravitational waves (GW). In addition to these known effects, quantum sensor networks~\cite{budker2015data} could be used to search for astrophysical signals predicted by beyond-standard-model (BSM) theories \cite{safronova2018search}. Exotic bosonic fields are ubiquitous features of BSM theories and appear while seeking to understand the nature of dark matter and dark energy and solve the hierarchy and strong CP problems. We consider the case where high-energy astrophysical events could produce intense bursts of exotic low-mass fields (ELFs). We propose to expand the toolbox of multi-messenger astronomy to include networks of precision quantum sensors that by design are shielded from or insensitive to conventional standard-model physics signals. We estimate ELF signal amplitudes, delays, rates, and distances of GW sources to which global networks of atomic magnetometers \cite{pospelov2013detecting,afach2018characterization} and atomic clocks \cite{derevianko2014hunting,wcislo2018new} could be sensitive. We find that, indeed, such precision quantum sensor networks can function as ELF telescopes to detect signals from sources generating ELF bursts of sufficient intensity. Thus ELFs, if they exist, could act as additional messengers for astrophysical events. 
}

Many of the great mysteries of modern physics suggest the existence of exotic fields with light quanta (masses $\ll {\rm 1~eV}$): the nature of dark matter \cite{Pre83,Abb83,Din83,Duf09,Gra15review} and dark energy \cite{Ark04,Fla09,Joy15}, the hierarchy problem \cite{Gra15}, the strong {\emph{CP}} problem \cite{Pec77a,Pec77b,Wei78,Wil78,Din81,Shi80,Kim79}, and the quest for a quantum theory of gravity \cite{Bai87,Svr06,Arv10}. Intense bursts of such ELFs could be generated by cataclysmic astrophysical events such as black hole or neutron star mergers \cite{bini2017deviation,baumann2019probing}, supernovae~\cite{Raf88,Raf99}, or other phenomena, such as the processes that produce fast radio bursts~\cite{Iwa15,Tka15}. Due to the small masses of the ultralight bosons being considered as possible ELFs, a high energy event is generally not required for ELF production. However, because of the feeble couplings of ELFs to standard model particles and fields, the ELF flux needs to be considerable in order for ELF signals to be detectable in experiments, especially in the case of ELFs from distant astrophysical sources. In particular, the high energies and unknown physics of binary black hole (BBH) mergers~\cite{Loeb2018-BH-signularities} leave open many interesting theoretical possibilities for ELF production. Remarkably, BBH mergers have already offered surprises: for example, GW150914 and GW170104 have been associated with unexpected gamma-ray emission~\cite{Verrecchia2017,Connaughton2016-FERMI}.


Quantum sensors~\cite{Degen2017-RMP-quantum-sensing} such as atomic clocks and magnetometers are sensitive to gentle perturbations of internal degrees of freedom (energy levels, spins, etc.) by coherent, classical waves. This is in contrast to particle detectors such as those employed in observations of cosmic neutrinos~\cite{halzen2017high}, gamma rays~\cite{holder2006first,atwood2009large}, and searches for weakly interacting massive particles (WIMPs) \cite{agnese2015improved,aprile2017first}. The key point is that in order to be detectable by the quantum sensors considered in the present work, the astrophysical source must produce coherent ELF waves with high mode occupation number. For example, if an axion burst resulted in just a few axions reaching the Earth, the effects would not be detectable with clocks and magnetometers. Thus we focus our attention on coherent production mechanisms for ELFs~\cite{bini2017deviation,arvanitaki2011exploring,hardy2017stellar,baumann2019probing} rather than thermal (incoherent) production mechanisms~\cite{Raf88,Raf99}.

There are several possibilities for the ELF production.
Could a black hole merger produce a transient burst of energy in the form of an ELF that is observable to the outside world through the vibrations it induces in the event horizon \cite{Loeb2018-BH-signularities}? Much of the underlying physics of coalescing singularities in black hole mergers remains unexplored as it requires understanding of the as yet unknown theory of quantum gravity~\cite{Loeb2018-BH-signularities}.   In addition, exotic scalar fields appear in theories that do not require invoking quantum gravity per se.  For example, rotating black holes may be surrounded by dense clouds of exotic bosons (with up to 10\% of black hole mass extracted by the clouds) that could lead to ELF bursts coincident with 
GW emission~\cite{arvanitaki2015discovering,arvanitaki2017black,baryakhtar2017black,baumann2019probing,yoshino2015probing}. Scalar fields also appear in well-posed theories of scalar-tensor gravity~\cite{Fujii:2003pa,Faraoni:2004pi,Deffayet:2013lga} resulting in black holes and neutron stars being immersed in scalar fields. Modes of these fields can be excited during BBH or binary neutron star (BNS) mergers~\cite{Franciolini2019}.  
Scalar  emission can be substantially enhanced due to dynamic scalarization~\cite{Barausse2013} and by the fact that the  scalar  emission is monopole in character~\cite{Krause1994a}.
Scalar fields can be trapped gravitationally in neutron stars~\cite{Garani2019} and can be potentially released during the BNS mergers. If  scalars are coupled to standard model particles and fields, they can be produced during BNS mergers. We refer the reader to a review on potential new physics signatures in GW events~\cite{Barack2019}. We also note that it has been proposed that there could be a direct coupling of spins to GWs \cite{bini2017deviation}, which would lead to a signal potentially detectable with atomic magnetometers. 

Considering the wide variety of speculative scenarios for ELF emission, here we take a pragmatic observational approach based on energy arguments.  GW events can radiate great amounts of energy, a fraction of which could be emitted in the form of ELFs. We assume that some amount $\Delta E$ of the total energy emitted by the astrophysical event is converted into  ELFs.  The radiated energy in the form of GWs from recently observed BBH mergers is a few solar masses ($M_\odot c^2$) \cite{abbott2016observation,abbott2017gw170814}, whereas for recently observed BNS mergers the radiated energy in the form of GWs is $\gtrsim 0.025 M_\odot c^2$ \cite{abbott2017gw170817}, where only a lower bound on energy release is obtained due to uncertainty about the equation of state for the neutron stars. For the purposes of the following sensitivity estimates, we assume that it may be possible to have $\Delta E \sim M_\odot c^2$ of energy released in the form of ELFs from a black hole merger and $\Delta E \sim 0.1 \,M_\odot c^2$ of energy released in the form of ELFs from a BNS merger.

\begin{figure*}[ht!]

\includegraphics[width=0.75\textwidth]{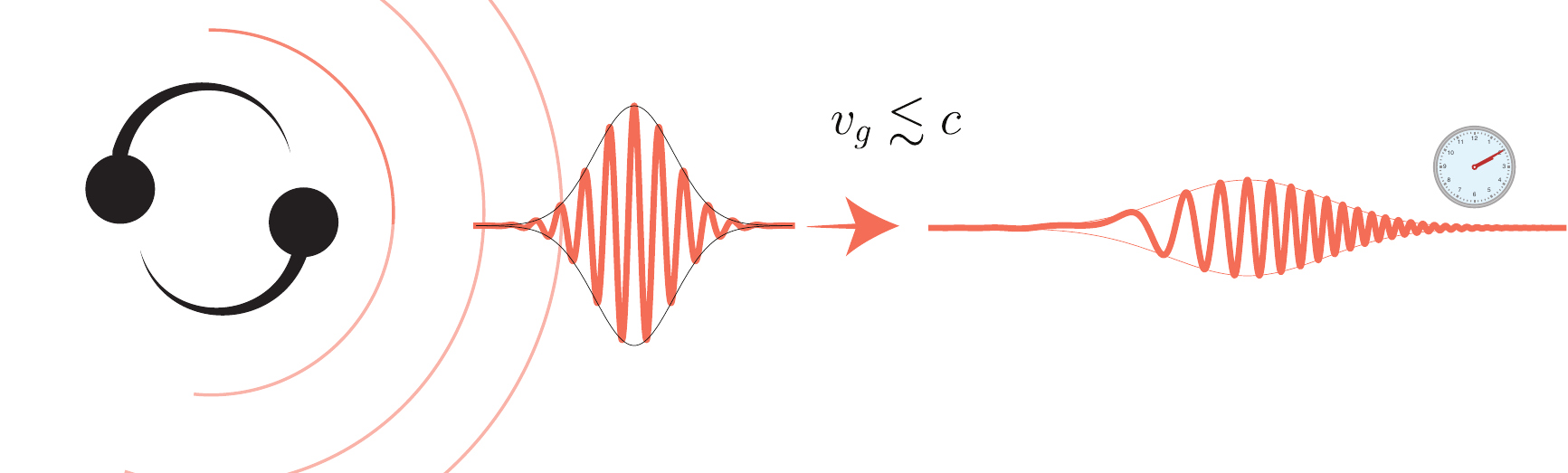}
\caption{Effect of dispersion on the expected ELF signal at  a precision quantum sensor. As the ELF burst propagates with  the group velocity $v_g \lesssim c$ to the detector,  it lags behind the GW burst.  Since the more energetic ELF components propagate faster, the arriving ELF wavepacket exhibits  a characteristic  frequency chirp. }
\label{Fig:Cartoon}
\end{figure*}

\begin{figure}[ht!]
\centering
\includegraphics[width=\columnwidth]{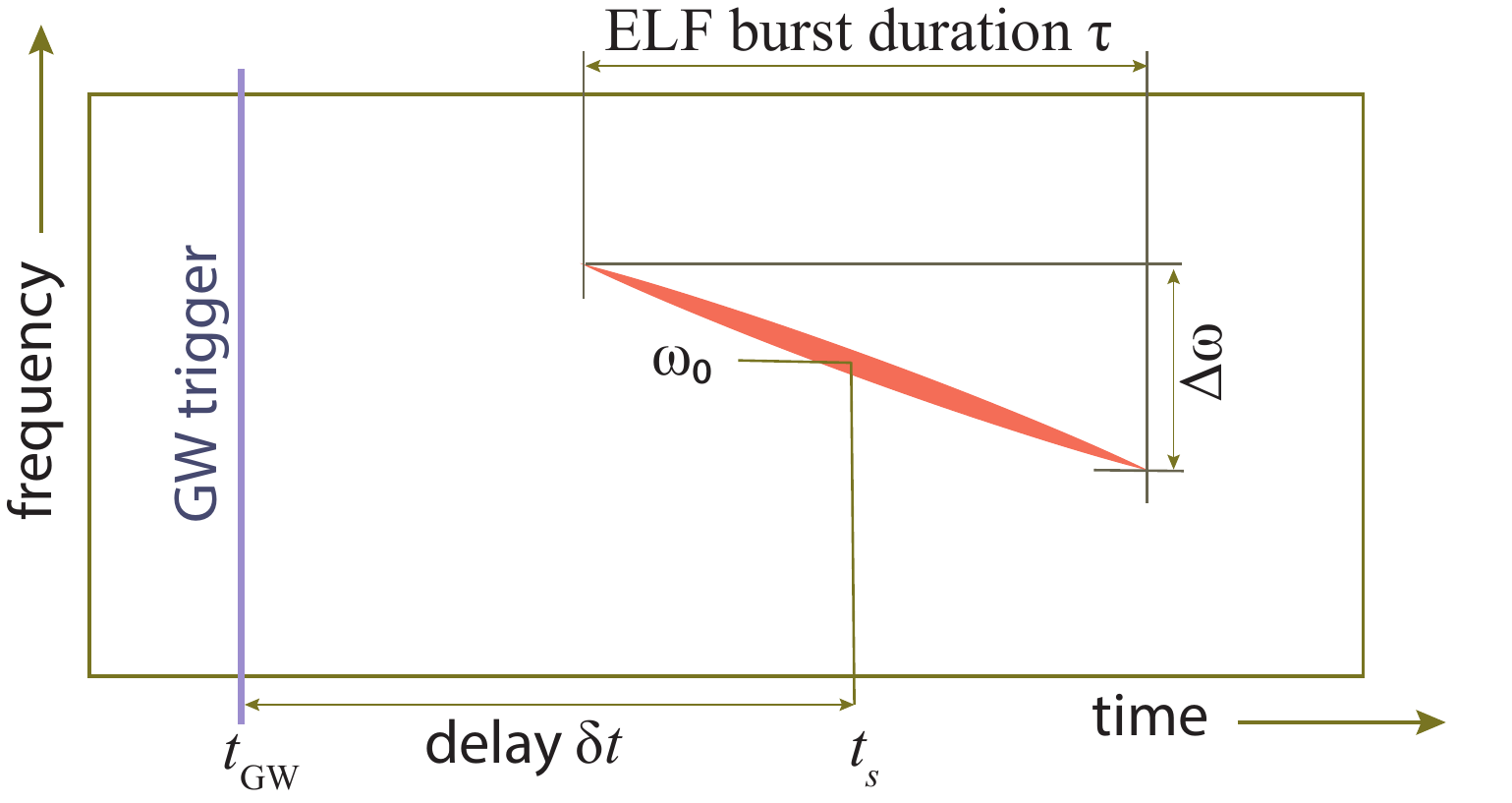}
\caption{ Time-frequency decomposition for an ELF signal for  interactions with sensor that is linear in the ELF field. The spectral width of the pulse $\Delta \omega$ is related to the initial pulse duration $\tau_0$ as $\Delta \omega =1/\tau_0$. The frequency slope is $d \omega(t)/dt = - \Delta \omega/\tau$. For  interactions quadratic in the ELF fields, the central frequency $\omega_0$ and the slope are doubled. 
 }
\label{Fig:ELF-template}
\end{figure}


For concreteness, we assume that the emitted ELF is a spin-0 field $\phi(r,t)$ described by a superposition of spherically symmetric wave solutions to the Klein-Gordon equation:
$\phi_k(r,t) = \frac{A_k}{r}\cos\prn{ k  r - \omega t  + \theta_k},$  
where $r$ is the radial coordinate, $A_k$, $\theta_k$, $k$, and $\omega$ are the ELF amplitudes, phases, wavevectors, and frequencies, respectively. The spherically symmetric monopole emission pattern is characteristic of scalar-tensor gravity models~\cite{Krause1994a}. 

The ELF frequency $\omega$  and  wavevector $k$
satisfy the relativistic energy-momentum dispersion relation,
$\omega(k)=\sqrt{(c k)^2+\Omega_c^2},$ 
where the  Compton frequency $\Omega_c = mc^2 /\hbar$ depends on  the ELF mass $m$. We consider ELFs sufficiently far from the source that general relativistic effects (such as the gravitational redshift) can be ignored. 
We  also ignore the effects of galactic dust~\cite{HensleyBull2018_dust} on the propagation and attenuation of the ELF waves.  


We consider an emitted ELF  burst  of central frequency $\omega_0$ and  of  a finite duration $\tau_0$, i.e. of bandwidth $\Delta \omega  \approx 1/\tau_0$  or, equivalently, of characteristic energy $\varepsilon_0 =\hbar \omega_0$  and  width $\Delta \varepsilon$. We decompose the wavepacket into spherical waves.  
 Individual Fourier components propagate with different phase velocities as dictated by the dispersion relation.
Higher frequency components propagate faster and we qualitatively expect a  frequency-chirped ELF signal at the detector, as shown in Figs.~\ref{Fig:Cartoon} and~\ref{Fig:ELF-template}. The  slope of the chirp is  $d \omega/dt  \approx - \Delta \omega/\tau$, since due to energy conservation the frequency content of a wavepacket is preserved. This estimate is supported  by explicit computations in the \SupMat.

 
With $R$ being the distance  from the  astrophysical source to the sensor, the ELF-GW  time delay is $\delta t = (R/c)  ( c/v_g - 1)$. In this formula, the wavepacket propagates over  time $t_\mathrm{GW} = R/c$, which  is $\sim$ a billion  years for GW150914. Thus $t_\mathrm{GW}$ is  much larger  than any reasonably observable time delay in an experiment (say $\delta t< 1\, \mr{week}$).  Therefore, $( c/v_g - 1) \ll 1$, and so to be observed ELFs must be ultrarelativistic.   In this limit, the ELF  central frequency $\omega_0$ and wavevector   $k_0$ are  related  by photonic dispersion $\omega_0 \approx c k_0$. The bandwidth of  a quantum sensor fixes  measurable ELF frequencies.  For atomic  clocks 
$\omega_0/2\pi \lesssim 1 \, \mathrm{Hz}$,  for atomic magnetometers $\omega_0/2\pi \lesssim 100\, \mathrm{Hz}$, and 
for optical cavities  $\omega_0/2\pi \lesssim 10 \, \mathrm{kHz}$.
These  frequencies fix energies $\varepsilon_0 $  of detectable ELFs  to  below $10^{-14} \,  \mathrm{eV}$ for clocks and $10^{-10} \,  \mathrm{eV}$ for cavities. Since the  dominant fraction of these  energies is of kinetic nature, the  fields  are necessarily ultralight, $m c^2 \ll \varepsilon_0$.   Emitted ELFs are copious ($ \gtrsim 10^{70}$ for $\Delta E \sim 0.1 M_\odot c^2$ and $\omega_0 = 2\pi \times 10 \,\mathrm{kHz}$).  The resulting mode occupation numbers at the Earth are macroscopic and therefore ELFs would act as coherent classical fields at the sensors.

The time delay  of the  ELF signal  with  respect to the  GW burst  is described by
$\delta t = \frac{t_\mathrm{GW}}{2} \left( \Omega_\mr{c}/\omega_0 \right)^2.$
 As   $\delta t \ll t_\mathrm{GW}$,   the Compton frequency $\Omega_\mr{c} \ll \omega_0$, consistent with ELFs being ultrarelativistic. 
The duration  $\tau$ of the ELF pulse at the sensor can be estimated as  $\tau  \sim R  \Delta  v_g/c^2$,  where the spread in group  velocities  $\Delta  v_g/c \approx  \partial^2\omega/\partial k^2/\tau_0$.  This leads to a relation between the signal duration and time delay
 $\tau \approx  2  \delta t \,    /(\omega_0 \tau_0).$ 
Since our approximations hold for sufficiently sharp ELF spectra, $\omega_0 \tau_0 \gg 1$, we require $\tau \ll  \delta t$.

The  characteristic ELF amplitude  $A_{k_0}$  at the sensor can be estimated by  requiring that the total energy of  the   scalar wave stored in a shell of thickness $c \tau$ and  radius $R$  to be  equal to the total energy $\Delta E$, 
$A_{k_0} \approx \frac{1}{\omega_0} \sqrt{ \frac{c\Delta E}{2\pi\tau}} .$  
In contrast to dispersionless spherical waves, the  field amplitude at the sensor  $\phi(R,t)$ scales  as  $1/R^{3/2}$,  reflecting the additional pulse dispersion.

More detailed considerations (see \SupMat) yield the following approximate time dependence for an ELF signal at the sensor,
\begin{align}
\phi(t) \approx & \frac{1}{R} \prn{\frac{c\Delta E}{2\pi^{3/2} \omega_0^2 \tau}}^{1/2} \exp\prn{-\frac{(t-t_s)^2}{2 \tau^2}} \nonumber\\
                  & \times \cos\prn{ \omega_0 (t-t_s) - \frac{\omega_0}{4\delta t}(t-t_s)^2 }\,,
\label{Eq:DispersionModeldE}
\end{align}
where $t_s = t_\mathrm{GW} +  \delta t $ is the time of arrival of the center of the pulse (see Fig.~\ref{Fig:ELF-template}). Note that the ELF frequency is time-dependent, $\omega(t)  = \prn{ 1 - (t-t_s) /(2\delta t)}\omega_0$, exhibiting a  frequency ``chirp" at the sensor.  The waveform, Eq.~\eqref{Eq:DispersionModeldE}, is shown in Fig.~\ref{Fig:Cartoon} and its power-spectrum time-frequency decomposition is shown in  Fig.~\ref{Fig:ELF-template}. The slope of the chirp (the line in Fig.~\ref{Fig:ELF-template}) is given by $d\omega/dt = - 1/\prn{\tau \tau_0}=-\omega_0/\prn{2\delta t}$, consistent with the qualitative arguments presented above. Data analysis to search for ELFs can be carried out using the excess power statistic as discussed in the \SupMat.





ELFs can generate signals in quantum sensors via ``portals'' between the exotic fields and standard model particles and fields.  Portals are a phenomenological gauge-invariant collection of standard model operators coupled with operators from the ELF sector~\cite{safronova2018search}.
 We consider interaction Lagrangians that are linear, $\mathcal{L}^{(1)}$, and quadratic, $\mathcal{L}^{(2)}$, in the ELF $\phi$. For magnetometers 
$\mathcal{L}^{(1)}_\mr{mag} = f_l^{-1} J^\mu \partial_\mu \phi$, 
$ \mathcal{L}^{(2)}_\mr{mag} = f_q^{-2} J^\mu \partial_\mu \phi^2$ 
and for  clocks, cavities, interferometers, and gravimeters: 
$\mathcal{L}^{(1)}_\mr{clk} =  \sqrt{4 \pi}/ E_\mr{Pl} \prn{ -d_{m_e} m_e c^2   \bar{\psi_e} \psi_e + d_e F_{\mu \nu}^2 /4}  \phi$, 
$\mathcal{L}^{(2)}_\mr{clk} = \prn{ -m_e c^2   \bar{\psi_e} \psi_e/\Lambda_{m_e}^2 + F_{\mu \nu}^2 /(4 \Lambda_{\alpha}^2)} \phi^2.$
In these expressions, $J^\mu$ is  the axial-vector current for SM fermions, $\psi_e$ is the  electron bi-spinor,  $F_{\mu\nu}$ is the Faraday tensor, $E_\mr{Pl}$  is the Planck energy,  and $f_{l,q}, d_e, d_{m_e}, \Lambda_{m_e},\Lambda_{\alpha}$ are coupling constants. Quadratic interactions appear naturally for ELFs possessing either $Z_2$ or $U(1)$ intrinsic symmetries~\cite{KimBudEby18}. 

The $\mathcal{L}_\mr{mag}$ portals lead to fictitious effective magnetic fields that interact with atomic spins and thus are detectable with atomic magnetometers \cite{pospelov2013detecting}. The $\mathcal{L}_\mr{clk}$ portals effectively alter fundamental constants~\cite{derevianko2014hunting}, such  as the electron mass $m_e$ and the fine-structure constant $\alpha$.
Such portals can imprint measurable signals in atomic clocks~\cite{derevianko2014hunting}, cavities~\cite{Cavity.DM.2018} and atom interferometers~\cite{GeraciDerevianko2016-DM.AI}. The $\mathcal{L}_\mr{clk}$ portals also modify the Earth's gravitational potential and thus can be detectable with gravimeters~\cite{GeraciDerevianko2016-DM.AI}. 

ELFs interacting through any of the enumerated portals would drive frequency-chirped signals in quantum sensors (Figs.~\ref{Fig:Cartoon} and \ref{Fig:ELF-template}), provided the sensors have sufficient sensitivity and bandwidth.  The coupling strengths determine, for a given ELF intensity, the relative signal amplitude detected by the particular sensor. 
In the \SupMat ~we show   that  the sensors can detect ELF bursts as  long as the coupling constants satisfy 
\begin{eqnarray}
f_l & \lesssim & \frac{\hbar^{3/2}c}{3} \frac{\sqrt{N_s}}{\sigma_m(\Delta_t) \sqrt{\Delta_t}} \frac{\sqrt{\Delta E}}{R} \, ,\label{Eq:LinSpinConstraint} \\
f_q &  \lesssim & 0.3  \frac{\hbar c}{R} \prn{ \frac{N_s}{\Delta_t \tau } }^{1/4}
\prn{ \frac{\Delta E}{\sigma_m(\Delta_t)  \omega_0} }^{1/2}\, ,\label{Eq:QuadSpinConstraint} \\
d_X & \gtrsim & 2 \,\frac{E_\mr{Pl}}{|K_X| } 
\frac{\sigma_y(\Delta_t) }{\sqrt{N_s} } \prn{  \frac{\omega_0}{c} R} \prn{\frac{\Delta_t}{\hbar \Delta E}}^{1/2}  \, ,\label{Eq:ModuliConstraint} \\
\Lambda_X & \lesssim & 0.1 \, 
\prn{ \frac{\sqrt{N_s} |K_X| }{\sigma_y(\Delta_t) }}^{1/2}
 \prn{  \frac{c}{R\omega_0}} 
  \prn{\frac{\hbar^2 \Delta E^2} {\Delta_t \tau}}^{1/4}.\label{Eq:LambdaConstraint}
\end{eqnarray}
Here $K_X$ is the sensitivity coefficient to a variation in fundamental constant $X=\{m_e,\alpha, \ldots\}$,  $\Delta_t$ is the sensor  sampling time interval, $\sigma_m(\Delta_t)$ is the magnetometer Allan deviation over $\Delta_t$ (in units of energy), $\sigma_y(\Delta_t)$ is the dimensionless clock/interferometer Allan deviation for  fractional frequency excursions, and $N_s$ is the number of sensors.  

Astrophysical observations and laboratory experiments set constraints on the coupling strengths between ELFs and standard model particles and fields~\cite{safronova2018search}. 
Using the above sensitivity estimates, we
 find that the current generation of atomic clocks is sensitive to quadratic  portals $\mathcal{L}^{(2)}$ as the prior constraints on  such  interactions are much weaker than those for the linear portals $\mathcal{L}^{(1)}$.
\begin{figure}[ht!]
\includegraphics[width=\columnwidth]{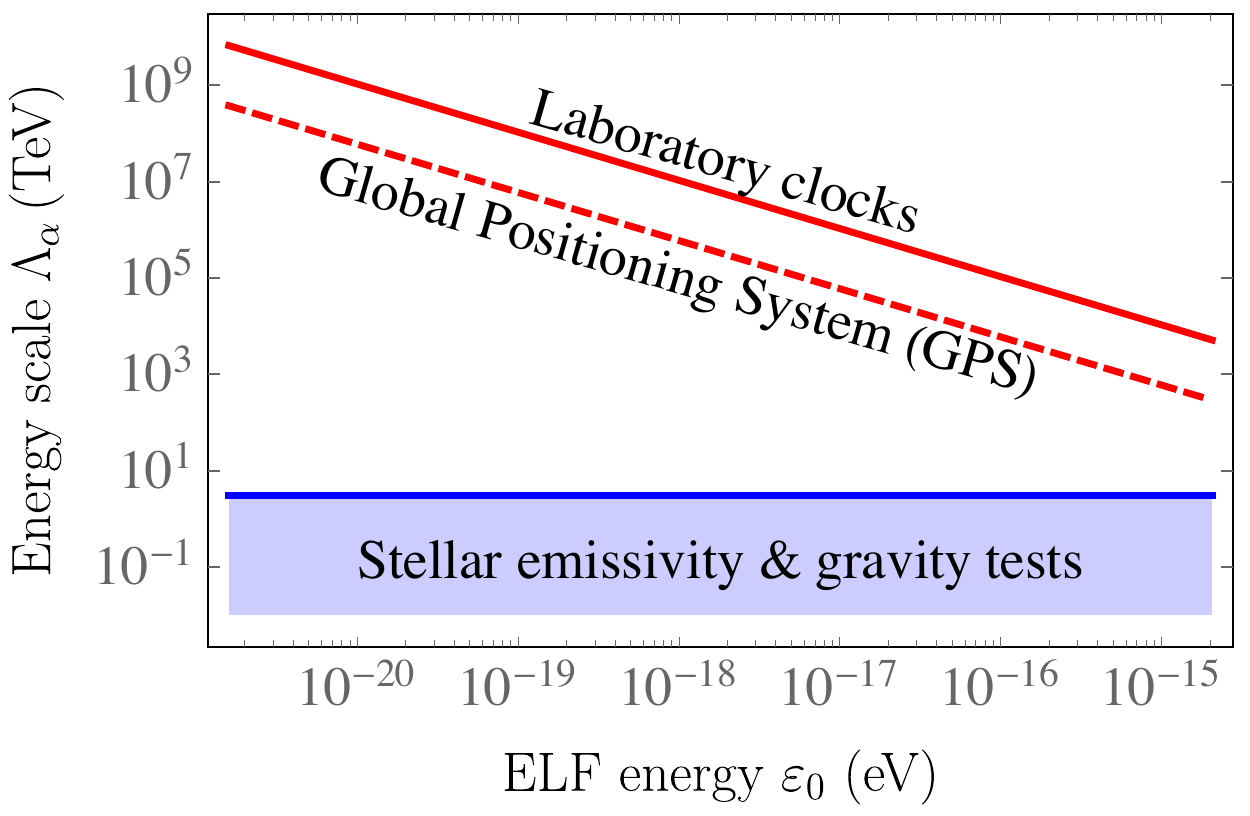}
\caption{Projected atomic clock sensitivity to ELFs plausibly emitted during the BNS merger GW170817.  
The discovery reach is shown for a trans-European network of  laboratory clocks (red line, $\sigma_y(1\, \mathrm{s}) =10^{-16}$)  and for the GPS constellation (red dashed line, $\sigma_y(1\, \mathrm{s}) =10^{-13}$).
We assumed an ELF burst of duration $\tau = 100 \,\mathrm{s}$, energy release $0.1 \Msun c^2$, and a total observation time of one month.  Prior constraints~\cite{Olive:2007aj} on the energy scale $\Lambda_\alpha$ are shown by the blue shaded region.
 }
\label{Fig:ClockLimits-quad}
\end{figure}

Several networks of precision quantum sensors are already operational. An example of an atomic clocks network is the Global Positioning system (GPS), nominally comprised of 32 satellites in medium-Earth orbit. The satellites house  microwave atomic clocks and they have been used for dark matter searches~\cite{roberts2017search,roberts2018search}. 
Combined with other satellite positioning constellations and terrestrial clocks, $N_s \sim 100$. Another network is  a trans-European fiber-linked network  ($N_s \sim 10$) of  laboratory clocks~\cite{Roberts2019-DM.EuropeanClockNetwork}  whose accuracy is vastly superior to the GPS clocks. 
As for magnetometers, the Global Network of Optical Magnetometers  for Exotic physics (GNOME) is a network of shielded optical atomic magnetometers with subpicotesla sensitivity. GNOME specifically targets transient events associated with beyond standard model physics~\cite{pospelov2013detecting,afach2018characterization,pustelny2013global,kimball2018searching,GNOMEwebsite}. Presently GNOME consists of $N_s = 12$  magnetometers located on three continents ~\cite{GNOMEwebsite}.

As an example, in Fig.~\ref{Fig:ClockLimits-quad}, we plot the projected sensitivity to a putative ELF burst emitted during the BNS  merger GW170817 ($R=40 \,\mr{Mpc}$). It is clear that existing clock networks can be sensitive to ELFs for a typical GW event (either BNS, BBH or BH+NS mergers) registered by GW detectors. If the sought ELF signal is not observed,
the sensors can place constraints on theoretical models.
The case of GPS is particularly intriguing as  $\sim 20$ years worth of archival GPS data is available and the dataset is routinely updated~\cite{MurphyJPL2015}.  If an ELF signal is discovered in  recent data, one can go back to pre-LIGO era  and search for similar signals in the archival data. Another  possibility is to correlate the catalogued short gamma ray bursts~\cite{Paul2017} or other powerful astrophysical events with the archival GPS data to search for  ELF bursts. Although estimates show that the existing magnetometer network does not have sufficient sensitivity to probe unconstrained parameter space for an ELF burst from GW170817 with the assumed characteristics, planned upgrades will substantially increase GNOME's discovery reach, as discussed in the \SupMat. 

Employing networks is crucial for distinguishing ELF signals from spurious noise. Furthermore, by having baselines with the diameter of the Earth or larger, one can resolve the sky position of the ELF source. This is a critical feature for multi-messenger astronomy that enables correlation with other observations of the progenitor. The leading edge of an ultrarelativistic ELF burst would propagate across the Earth in $\sim 40~{\rm ms}$. GNOME magnetometers presently have a temporal resolution of $\approx 10~{\rm ms}$, this can be  improved to $\lesssim 1 \, \mr{ms}$ with relatively straightforward upgrades \cite{budker2013optical}. The angular resolution $\Delta\theta$ based on the ELF time-domain signal pattern is given roughly by the ratio of the temporal resolution to the propagation time through the network:  for a temporal resolution of $\approx 1~{\rm ms}$ this corresponds to $\Delta\theta \approx \pi/40~{\rm rad} \approx 2^\circ$. Additionally, since the ELF gradient points along the ELF velocity vector, the relative signal amplitudes in magnetometers with different sensitive axes enables a second method of angular resolution of the source's sky position. The signal amplitude pattern in the network would yield an angular resolution (in radians) roughly equal to the inverse of the signal-to-noise ratio for the ELF detection.

Unlike magnetometers, atomic clocks and atom interferometers have a relatively low $\sim 1 \, \mathrm{Hz}$ sampling rate. As a result, terrestrial or satellite clock networks cannot be used to track the ELF burst propagation. The ELF propagation time across the GPS constellation is $0.2 \, \mr{s}$,
which is comparable to the $1 \, \mr{s}$ sampling interval in GPS datastreams.
Nonetheless, clock networks can still act collectively, gaining $\sqrt{N_s}$ in sensitivity and vetoing signals that do not affect all the sensors in the network. To mitigate the low sampling rate, one can envision increasing the baseline, similar to recently proposed~\cite{Tino-SAGE-2019} space-based GW detectors relying on atomic clocks and atom interferometers. Another possibility is a small-scale ($\sim 10 \, \mathrm{km}$) terrestrial network of optical cavities which allow for $ \gtrsim 10\,  \mr{kHz}$ sampling rate. Each node of  such a network would contain two cavities~\cite{Cavity.DM.2018}: one with a rigid spacer and the other with suspended mirrors. An ELF-induced variation in fundamental constants would change the length and thus the resonance frequency of the former while not affecting that of the latter. 
The ELF sensitivity of a cavity network is similar to that of the clock networks shown in  Fig.~\ref{Fig:ClockLimits-quad}.

In conclusion, we have demonstrated the ability of global networks of precision quantum sensors to detect exotic low-mass fields (ELFs) that can be plausibly emitted from high energy astrophysical events, potentially making ELFs new messenger modality in the growing field of multi-messenger astronomy.

\section*{Acknowledgments}
\label{Sec:Acknowledgments}

We  thank L. Bernard,  G. Blewitt, S. Bonazzola, D. Budker,   A. Furniss, S. Gardner, E. Gourgoulhon, K. Grimm, M. Pospelov, J. Pradler, B. Safdi,  J. E. Stalnaker, and  C. Will for discussions. This work was supportted in part by  the European Research Council (ERC) under the European Unions Horizon 2020 research and innovation program (grant agreement No 695405), the DFG Reinhart Koselleck project, the Simons and Heising-Simons Foundations, and from the U.S. National Science Foundation under Grant No.~PHY-1707875,  PHY-1806672, and PHY-1912465.

\newpage
\section{ }\newpage
\section{Supplementary Information}

\section{ Energy density for spherical wave of ultrarelativistic scalar field}
\label{Appendix:Amplitude}
In the main text we expand the real-valued scalar field in spherical waves,
\begin{align}
\phi_k(r,t) = \frac{A_k}{r}\cos\prn{ k  r - \omega t  + \theta_k}\,.
\label{Eq:App:spherical-wave-scalar}
\end{align}
Here $r$ is the radial coordinate, $A_k$ and $\theta_k$ are the ELF amplitudes and phases 
and $k$ and $\omega$ are ELF wavenumber and oscillation frequency.
The  field $\phi_k$ has units of $M^{1/2} L^{1/2} T^{-1}$ and the amplitude $A_k$ has the units of $M^{1/2} L^{3/2} T^{-1}$.

The energy density $\rho$ is given by the $00$ component of the stress-energy tensor~\cite{PeskinSchroeder1995QFTbook}, 
\begin{equation}
\rho  = \frac{1}{2c^2}\dot{\phi}^2 + \frac{1}{2}( \mb{\nabla} \phi )^2 + \frac{1}{2}\frac{m^2c^2}{\hbar^2}\phi^2 \,,
\label{App:Eq:EnergyDensity}
\end{equation}
where $m$ is the mass of the scalar.
Explicitly,  for a spherical wave~(\ref{Eq:App:spherical-wave-scalar}),
\begin{align}
\rho =\frac{A_k^2}{2r^2}\frac{\omega^2}{c^2} \Bigg[\sin^2( \cdots) + \prn{\frac{ck}{\omega}}^2\sin^2(\cdots)\nonumber\\
+\prn{\frac{mc^2}{\hbar\omega}}^2\cos^2(\cdots)\Bigg] + \text{O}\prn{\frac{1}{r^3}}\,,
\end{align}
where $\cdots$ stands for the argument of  cosine  in Eq.~(\ref{Eq:App:spherical-wave-scalar}). We  neglect terms of order $1/r^3$, take the time average over many field oscillations,  and employ the ultrarelativistic limit, $\omega \approx ck \gg mc^2/\hbar$. The resulting energy density reads
\begin{align}
\abrk{\rho} \approx \frac{1}{2}\prn{\frac{A_k}{r}\frac{\omega}{c}}^2\,.
\end{align}

\section{Dispersion of ultra-relativistic matter wave pulse}
\label{App:Dispersion}

Any type of wave will disperse upon propagation as long as the dispersion relation $\omega(k)$ has a nonzero second derivative with respect to $k$. This ensures that the group velocity is a function of $k$. Here we focus on an analytically tractable case of a Gaussian wavepacket composed of ultrarelativistic scalar fields.

{\em Dispersion relation in the ultrarelativistic limit ---} 
We start  with the Klein-Gordon equation for the scalar field $\phi(\mathbf{r},t)$, $ (\partial_\mu\partial^{\mu} + m^2c^2/\hbar^2) \phi(\mathbf{r},t) =0$.
Focusing  on the spherically-symmetric solutions  ($s$-waves, characteristic of scalar emission in scalar-tensor theories), we define  $\phi(r,t)=u(r,t)/r$. Then the Klein-Gordon equation reduces to the 1D wave equation for massive scalar fields
\begin{equation}
  \left( 
  \frac{1}{c^2}\frac{\partial^2}{ \partial t^2}  - \frac{\partial^2}{ \partial r^2} + \frac{m^2c^2}{\hbar^2} \right) u(r,t) = 0\,.
\end{equation}
Substitution of $u(r,t) \propto \exp( i k r  \pm i \omega  t )$  leads, as expected,  to  the  relativistic energy-momentum relation 
\begin{equation}
 \omega(k) = \sqrt{ (ck)^2  + (mc^2/\hbar)^2}\, ,\label{App:Eq:dispersion}
 \end{equation}
i.e., the dispersion relation in the main text. Of course, it holds for waves of arbitrary angular momentum. We recognise here that $\omega_0=\omega(k_0)$. In the ultrarelativistic limit, $ck \gg mc^2/\hbar$, the energy of an individual scalar $\varepsilon \approx c|k|$.

We can further expand  $\omega(k)$  around a  characteristic  energy $\varepsilon_0 = \hbar \omega_0\approx c\hbar k_0$,
\begin{equation}
\omega(k) \approx \omega_0 + c\frac{ck_0}{\omega_0} (k-k_0) + \frac{1}{2}\prn{\frac{mc^2}{\hbar\omega_0}}^2\frac{c^2}{\omega_0}  (k-k_0)^2   \,,
\label{App:Eq:ParabolicDispersion}
\end{equation}
where we keep terms up to second order only. This  parabolic approximation  holds as long as $ |k -k_0|  \ll k_0$ or, equivalently, when  the energy spectrum of  emitted scalars is sufficiently sharp,  $\Delta \varepsilon \ll \varepsilon_0$, or $\omega_0 \tau_0 \gg 1$.  One  can immediately identify the group velocity
\begin{equation}
\frac{v_g}{c} = \frac{ck_0}{\omega_0}\approx\, 1-  \frac{1}{2}  \left( \frac{m c^2}{\varepsilon_0} \right)^2  \, ,   \label{App:Eq:group}
\end{equation}
and the characteristic spread  in   group velocities 
\begin{equation}
\frac{\Delta v_g}{c} = \left( \frac{m c^2}{\varepsilon_0} \right)^2  \frac{ \Delta \varepsilon}{\varepsilon_0}\, ,
\label{App:Eq:deltaGroup}
\end{equation}
where $\Delta\varepsilon = \hbar/\tau_0$. Finally, the time lag between gravitational wave (GW) and ELF bursts  at the sensor a distance $R$ away from the progenitor is
\begin{equation}
\delta t = \left( \frac{m c^2}{\varepsilon_0} \right)^2  \frac{R}{2c}. 
\label{App:Eq:Lag}
\end{equation}
Eqs.~(\ref{App:Eq:group}--\ref{App:Eq:Lag}) are the  relations  used in the  main body of the paper. 

\begin{figure}[ht!]
\includegraphics[width=0.9\columnwidth]{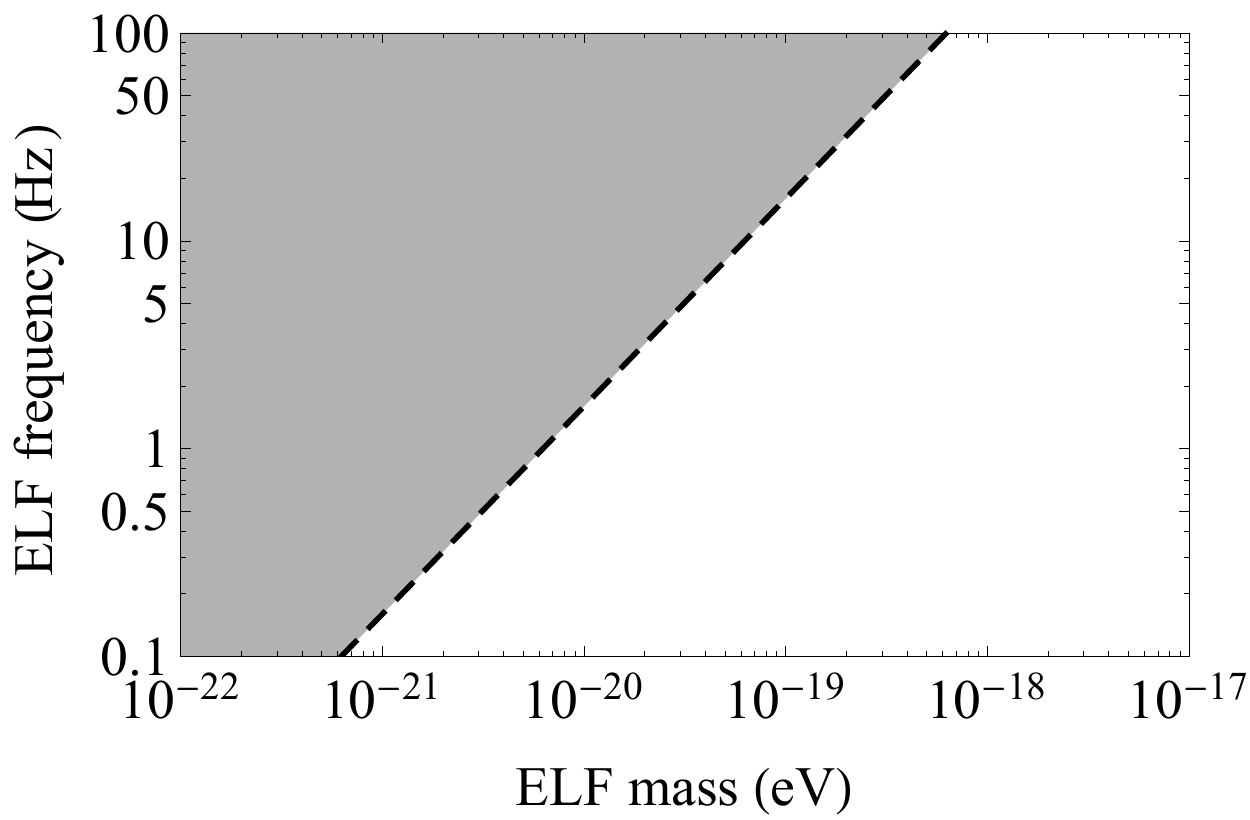}
\caption{Accessible parameter space (grey shaded region bounded by dashed line) for ELF detection with a network of quantum sensors based on the requirement that the maximum observed delay $\delta t$ [Eq.~\eqref{App:Eq:Lag}] of the ELF burst is $\lesssim 10$~hours.  Astrophysical parameters are taken for the GW170608 BBH coalescence event~\cite{abbott2017gw170608}: the distance to the source $R \approx 10^9$~ly and the characteristic  duration at the source is $\tau_0 \approx 1$~s. The ELF frequency $\omega_0/(2\pi)$ corresponds to the center frequency of the chirped pulse observed with the quantum sensor network.}
\label{Fig:accessibleMF}
\end{figure}

To illustrate the effect of the delay on the detectable ELF mass $m$, Fig.~\ref{Fig:accessibleMF} shows the accessible parameter space for an ELF burst associated with the GW170608 BBH coalescence event~\cite{abbott2017gw170608} assuming that the delay $\delta t$ is less than 10~hours. 

The  general solution to the 1D wave equation is a superposition of waves weighted by Fourier amplitudes $a(k)$,
\begin{equation}
u(r,t) = \frac{1}{\sqrt{2\pi}}\text{Re}\sbrk{\int^\infty_{-\infty} a(k) e^{i\prn{k r-\omega(k) t}} dk}~,
\label{Eq:GeneralKG}
\end{equation}
with the dispersion relation~(\ref{App:Eq:dispersion}). The initial conditions define the Fourier amplitudes~\cite{JacksonEM}
\begin{equation}
a(k) = \frac{1}{\sqrt{2\pi}}\int^\infty_{0} e^{-i k r}\sbrk{u(r,0)+\frac{i}{\omega(k)}\pdbydt{u}(r,0)}dr\,,
\label{App:Eq:IFT}
\end{equation}
with $u(r,0)$ and $\partial u/\partial t(r,0)$ being the initial values near the source.

{ \em Propagation and dispersion of a Gaussian wave packet --- }
Specializing our discussion to a {\em Gaussian} wave packet~\cite{JacksonEM} with initial wave amplitude $A_0$, initial spatial width $L_0$, and initial wavevector $k_0$:
\begin{align}
u(r,0) & ~ = ~ A_0 e^{-r^2/(2L_0^2)}\cos \prn{k_0 r}~, \nonumber\\ 
\pdbydt{u}(r,0) & ~ = ~ 0~.
\end{align}

The outgoing wavepacket has the Fourier amplitude
\begin{equation}
a(k) = \frac{A_0 L_0}{2}e^{-(L_0^2/2)(k-k_0)^2} \, ,
\label{Eq:FourierAmp}
\end{equation}
which implies the well-known uncertainty relation between the characteristic spatial extent $L_0$ of the wavepacket and its width in momentum space  $\Delta k \sim 1/L_0$.   Substitution of the above Fourier amplitude into 
Eq.~(\ref{Eq:GeneralKG}) fully solves the problem of propagation. We will use the parabolic approximation~(\ref{App:Eq:ParabolicDispersion}) for the dispersion  relation, which holds as long as $ \Delta k \ll k_0$, i.e.,  the characteristic wavelength of the field is much smaller than the initial spatial width $L_0$.  The parabolic dispersion allows the integral~(\ref{Eq:GeneralKG}) to be evaluated in a closed form. 

The final solution for $\phi(r,t)$ reads
\begin{equation}
\phi(r,t) \approx \frac{A_0}{r}\sqrt{\frac{\tau_0}{\tau(t)}}\exp \prn{-\frac{(t-r/v_g)^2}{2\tau(t)^2}}\cos \prn{\theta(r,t)} \, , \label{App:Eq:GaussianWavePacketFinal}
\end{equation}
with time-dependent pulse duration $\tau(t)$ defined as
\begin{equation}
\tau(t) = \sqrt{\tau_0^2+\prn{\frac{\Delta v_g t}{v_g}}^2}~,
\end{equation}
and we have substituted $L_0/v_g=\tau_0$.  The phase argument of the oscillatory part is given by
\begin{align}\label{Eq:phase}
\theta(r,t) = &~(\omega_0 t-k_0 r)-\frac{1}{2 \tau(t)^2}\frac{\Delta v_g t}{v_g\tau_0}\prn{t-r/v_g}^2\nonumber\\
&+\frac{1}{2}\tan^{-1} \prn{\frac{\Delta v_g t}{v_g \tau_0}}~.
\end{align}
In these expressions, group velocity $v_g$ and its spread $\Delta v_g$  are given by Eqs.~(\ref{App:Eq:group}) and (\ref{App:Eq:deltaGroup}).  Focusing on the sensor $(t  = t_s \equiv R/v_g )$, we define  the combination 
\[
 \xi =   \frac{\Delta v_g}{v_g} \frac{t_s}{\tau_0} = 2 \frac{\delta t} {\tau_0}  \frac{\Delta \varepsilon}{\varepsilon_0}  ,
\]
where $\delta t$ is the time lag~(\ref{App:Eq:Lag}) between the arrivals of GW  and  ELF bursts. 
When $\xi  \ll 1$, the  duration of the signal  at the detector
\begin{equation}
\tau \approx \frac{\Delta v_g}{v_g} t_s  =  2  \frac{\Delta \varepsilon}{\varepsilon_0}  \delta t \, .
\label{App:Eq:duration}
\end{equation}

Another important feature of the analytical waveform (\ref{App:Eq:GaussianWavePacketFinal}) is that it has an amplitude that scales  as $\tau(t)^{-1/2}$, as expected from the total energy conservation arguments of the main text. 
To relate the  amplitude $A_0$ to the total energy released in the ELF channel $\Delta E$, we compute the energy density $\rho(r,t)$, Eq.~(\ref{App:Eq:EnergyDensity}), for the Gaussian wavepacket (\ref{App:Eq:GaussianWavePacketFinal}). In the ultrarelativistic limit, 
\begin{equation}
    \rho(r,t)\approx\frac{1}{2c^2}\dot\phi^2+\frac{1}{2}\prn{\frac{\partial\phi}{\partial r}}^2 \, .
\end{equation}
While evaluating the derivatives of the  field it is sufficient to  keep the derivatives of the rapidly oscillating $\cos(\theta(r,t))$ factor. Then at a fixed time, we evaluate the pulse energy by integrating energy density over the space, leading to a time-independent value as expected. From here we  express the amplitude $A_0$ in terms of the total energy,
\begin{equation}
    A_0  \approx \frac{1}{\pi^{1/4}}\prn{\frac{1}{\omega_0}\sqrt{\frac{c\Delta E}{2\pi \tau_0}}}\, .
    \label{App:Eq:AmplitudeGaussian}
\end{equation}
 
{\em ELF  signal at the sensor -- } 
We define the instantaneous frequency 
$\omega(t) = d\theta(R,t)/dt$ and  expand it around the time the center of the pulse arrives at Earth, $t_s=R/v_g$.
\begin{align}
\omega(t) & \approx \omega(t_s) - \frac{d\omega}{dt}\Big|_{t_s}(t-t_s) \nonumber\\
          & \approx \omega_0 - \frac{1}{\tau_0 \tau}(t-t_s)~.
\end{align}
The sign of the linear term is consistent with the qualitative expectation of higher frequencies arriving first, and lower ones last. The slope  of the frequency chirp is given by 
\begin{equation}
\frac{d\omega(t)}{dt} = - \frac{1}{\tau_0 \tau} = - \frac{\omega_0}{2 \delta t} \,.
\label{App:Eq:Slope}
\end{equation} 
Then at the sensor,  the Gaussian ELF burst has an approximate temporal waveform,
\begin{align}
\phi(t) \approx & \frac{A_0}{R}\sqrt{\frac{\tau_0}{\tau}}\exp\prn{-\frac{(t-t_s)^2}{2 \tau^2}} \nonumber\\
                  & \times \cos\prn{ \omega_0 (t-t_s) - \frac{\omega_0}{4\delta t}(t-t_s)^2 }\,,
\label{App:Eq:DispersionModel}
\end{align}
 or, with Eq.~(\ref{App:Eq:AmplitudeGaussian}) for the amplitude, 
 \begin{align}
\phi(t) \approx & \frac{1}{R} \prn{\frac{c\Delta E}{2\pi^{3/2} \omega_0^2 \tau}}^{1/2} \exp\prn{-\frac{(t-t_s)^2}{2 \tau^2}} \nonumber\\
                  & \times \cos\prn{ \omega_0 (t-t_s) - \frac{\omega_0}{4\delta t}(t-t_s)^2 }\,.
\label{App:Eq:DispersionModeldE}
\end{align}

 {\em General envelope --- }
 The preceding analytical results explicitly demonstrate propagation and dispersion of a Gaussian wavepacket. These results  hold for a much wider class of sufficiently well-behaved envelopes. Formally, this can be shown by applying the stationary phase method while evaluating the integral~(\ref{Eq:GeneralKG}) for the parabolic dispersion relation~(\ref{App:Eq:ParabolicDispersion}). The stationary phase method effectively reduces the wavepacket to a Gaussian and all the derived results immediately apply.
 
\section{Data analysis considerations}
\label{App:Sec:DataAnalysis}
The goal of this section is  to outline a  data  analysis strategy and to establish  projected  sensitivity of the proposed search for a generic ELF signal.  To  reiterate,
an arriving ELF wavepacket can be characterized by a set of three parameters 
\begin{equation}
(\delta t, \tau,\omega_0)\,,
\end{equation} 
i.e., by the GW-ELF time delay $\delta t$, duration $\tau$, and central frequency $\omega_0$ (see Fig.~2 of the main text). Notice that the frequency chirp of the pulse is fixed by these parameters through Eq.~(\ref{App:Eq:Slope}).    
Since our approximations hold for sufficiently sharp ELF spectra, $\omega_0 \tau_0 \gg 1$ (see Sec.~\ref{App:Dispersion}), from  Eq.~(\ref{App:Eq:duration}), we expect $\tau \ll  \delta t$.

\begin{figure}[ht!]
\includegraphics[width=0.9\columnwidth]{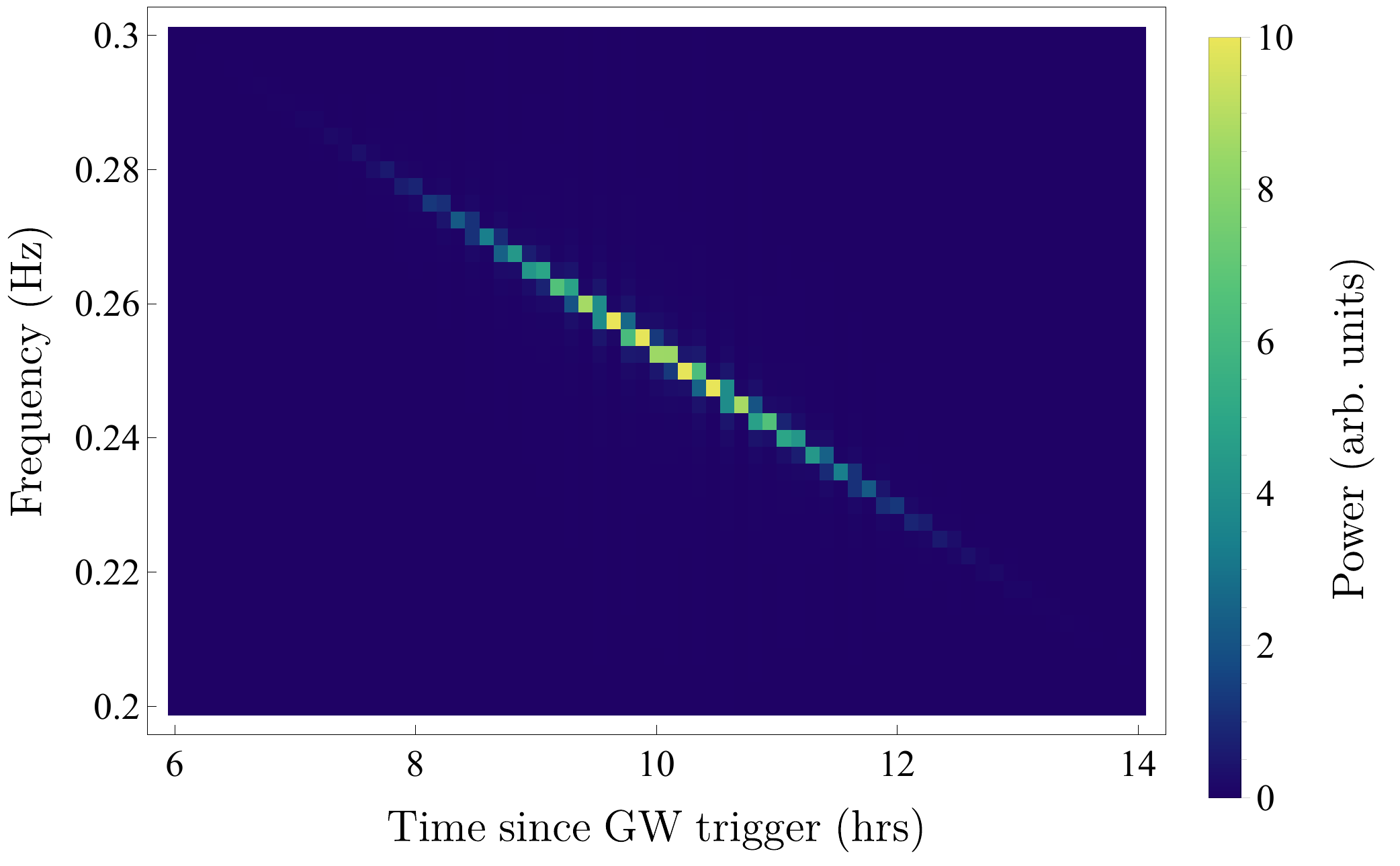}
\caption{A simulated ELF signal in time-frequency space (cf. Fig.~2 of the main text). The signal is computed for the Gaussian waveform  \eqref{App:Eq:DispersionModeldE}, with a central frequency  $f_0=\omega_0/(2\pi)=0.25$~Hz,  time delay $\delta t=10$~hours, 
initial pulse duration $\tau_0=10\,\mathrm{s}$, and  distance to the progenitor $R=40$~Mpc.
}
\label{App:Fig:dispersionExample}
\end{figure}

Given the parameters $(\delta t, \tau,\omega_0)$  and the known GW travel time from the progenitor $t_\mathrm{GW}=R/c$, one can fully determine other parameters. In particular, the
ELF particle mass (cf. Eq.~(\ref{App:Eq:Lag}))
\begin{equation}
 m = \frac{\hbar \omega_0}{c^2} \sqrt{ \frac{2 \delta t}{t_\mathrm{GW} } } \, ,
  \label{App:Eq:mass}
\end{equation}
and the initial pulse duration
\begin{equation}
 \tau_0 = \frac{2}{\omega_0 \tau} \delta t\, .
 \label{App:Eq:tau0}
\end{equation}
For a fixed total energy $\Delta E$ released into the ELF channel, the maximum field amplitude at the sensor is fixed to
\begin{equation}
\phi_{\mathrm{max}} \approx \frac{1}{R} \prn{\frac{c\Delta E}{2\pi^{3/2} \omega_0^2 \tau}}^{1/2} \,,
\label{App:Eq:PhiMax}
\end{equation}
where we take the amplitude for the Gaussian envelope~(\ref{App:Eq:DispersionModeldE}) as a fiducial value.

Considering a variety of  ELF production scenarios, we leave the envelope of the arriving wavepacket undefined. This uncertainty can be incorporated  into statistical analysis using the excess power statistic~\cite{Anderson2001}.
 This method is based on the time-frequency decomposition of the data, and detects events based on their signature of having more power in a time-frequency interval than one expects from detector noise alone. Excess power is the optimal method for searching for events in situations for which only a rough idea of the frequency and duration of the signal is known~\cite{Anderson2001, Maggiore}. 

Suppose the data streams from the sensors are sampled uniformly at a rate $1 / \Delta_t$ -- yielding a time series $\mathbf{d}$ with elements ${d_1, d_2 \ldots d_j\ldots d_{N_\mathrm{tot}}}$ for a data set with $N_\mathrm{tot}$ points.  Each data point $d_j = s_j + n_j $ comprises contributions from both the sought ELF signal, $s_j$, and intrinsic sensor noise, $n_j$.

Using the discrete Fourier transform (DFT) in  a sliding time window, the data stream can be partitioned into segments of time-and-frequency (tiles). Our goal is to quantify the power contained in each time-and-frequency tile of the data due to only noise, and thereby extract contributions due to putative ELF signals. To this end, the data stream can be split into two gross segments: before  and after the electromagnetic or GW triggers on detectors on Earth. The noise characteristics can be fully determined from the pre-trigger data, since during that period $d_j  =  n_j$ by our assumptions. We  assume that the sensor noise is Gaussian distributed and stationary but not necessarily  white (which, with appropriate filtering, is generally the case for the GNOME and GPS data \cite{afach2018characterization,Roberts2017-GPS-DM}).  Below we focus on a single sensor and later generalize to a network of  sensors.

The time series $\mathbf{d}$ is partitioned into segments containing $N_\mathrm{w}$ elements. $N_\mathrm{w}$  is chosen to be an even number for notational convenience. Each segment is associated with a data index $w$, coinciding with the mid-point of the partition: $w = N_\mr{w}/2, 3N_\mr{w}/2, 5N_\mr{w}/2 \ldots$, and a time $t_{w} =  {w} \Delta_t$.

The  Fourier amplitudes for each time partition are then given by
\begin{equation}
    \tilde{d}_{p,w} = \sum_{j=w-N_\mathrm{w}/2}^{w+N_\mathrm{w}/2} d_{w-j} e^{2\pi i (w-j) p /N_\mathrm{w}}\,,
    \label{App:Eq:Fourier}
\end{equation}
where index $p$ enumerates DFT frequencies, and $f_p=p/(N_\mathrm{w}\Delta_t)$ ranges from zero to the Nyquist frequency $1/(2\Delta_t)$. The DC ($f_p =0$) and Nyquist frequency amplitudes can be removed from the analysis since their statistical properties differ from the rest the amplitudes (see, e.g., Refs.~\cite{Derevianko2016a,RomanoCornish2017}). This simplification does not alter the conclusions. Eq.~(\ref{App:Eq:Fourier}) represents a 2D discrete map of complex time-frequency values. The frequency and time indices reference individual tiles $(p,w)$ in such a map.

Using the pre-event data  ($d_k \equiv n_k)$, we determine the (two-sided) power spectral density (PSD)  of the sensor noise
\begin{equation}
\tilde{C}_{p}\equiv \langle \tilde{n}_{p} \left(\tilde{n}_{p}\right) ^{\ast}\rangle \, ,
\end{equation}
where the averaging is over multiple pre-event time windows.  The
 post-event data PSD is normalized to the noise PSD
\begin{equation}
    \mathcal{E}_{p,w} \equiv \frac{| \tilde{d}_{p,w}|^2}{\tilde{C}_p} \, .
    \label{Eq:StatisticOneTile}
\end{equation}
The quantities $\mathcal{E}_{p,w}$ quantify excess power in the $(p,w)$ tile.\footnote{Note that our definition of excess power is larger by a factor  of  $2$ compared to Ref.~\cite{Anderson2001}.} In the absence of the sought-after ELF signal, $\langle \mathcal{E}_{p,w} \rangle =1$. A time-frequency decomposition map  for a Gaussian ELF wavepacket~(\ref{App:Eq:DispersionModeldE}) is shown in Fig.~\ref{App:Fig:dispersionExample}.


 We  adopt the method of  Ref.~\cite{Anderson2001} to incorporate our knowledge about the expected ELF signals. In that work, the search method probes all tiles occupying a rectangular area in the time-frequency decomposition map. Here, we restrict the probed tiles to the  ``fat line''  or ``scar'' areas spanned by the expected ELF signals.  Indeed,   the expected ELF  signal with the fixed parameter triple $(\delta t, \tau,\omega_0) $ contains significant power only in a subset of tiles, see Fig.~2 of the main text and Fig.~\ref{App:Fig:dispersionExample}.  Thereby, we define the excess power statistic $\mathcal{E}_\mathrm{ELF}$ by summing over the ELF-containing tiles
\begin{equation}
   \mathcal{E}  = \sum_{(p,w) \in \mathrm{ELF}}  \mathcal{E}_{p,w}  \, .
    \label{App:Eq:Statistic}
\end{equation}
We denote the total number of ELF-containing tiles as $M$. In the absence of noise in the post-event data, the total excess power contained in the ELF signal is
\begin{equation}
   \mathcal{E}_\mathrm{ELF}  = \sum_{(p,w) \in \mathrm{ELF}} \frac{| \tilde{s}_{p,w}|^2}{\tilde{C}_p}  \, .
    \label{App:Eq:StatisticELF}
\end{equation}

The probability distribution function for the statistic $\mathcal{E}$ is~\cite{Groth-1974}
\begin{align}
    &p_M(\mathcal{E}|\mathcal{E}_\mathrm{ELF} ) =    \label{App:Eq:chi2Non-central} \\
    &{I_{M-1}\prn{2\sqrt{\mathcal{E} \mathcal{E}_\mathrm{ELF} }}}\prn{\sqrt{\frac{\mathcal{E}}{ \mathcal{E}_\mathrm{ELF}}}\;}^{M-1} e^{- (\mathcal{E}+\mathcal{E}_\mathrm{ELF}) }\,,  \nonumber
\end{align}
where $I_{M-1}(\cdots)$ is the
modified Bessel function.  This distribution can be recognized, up to a change of scale, as a non-central $\chi^2$ distribution with $2M$ degrees of freedom. The mean and variance are given by
\begin{equation}
    \abrk{\mathcal{E}}=M+\mathcal{E}_\mathrm{ELF},\quad\text{Var}(\mathcal{E})=M+2\mathcal{E}_\mathrm{ELF} \,. \label{App:Eq:ExpVar}
\end{equation} 

Next  we would  like to establish the discovery reach for $\mathcal{E}_\mathrm{ELF}$ at the 95\% confidence level.
To  this end we  compute the upper tail probability threshold given the observed value  $\mathcal{E}_\mathrm{obs}$  of the statistic~(\ref{App:Eq:StatisticELF}) (the observed value is computed with sensor data)
\begin{equation}
    \int_{\mathcal{E}_{\mathrm{obs}}}^\infty p_M(\mathcal{E} |\mathcal{E}^{95\%}_\mathrm{ELF})\;d\mathcal{E} =0.95 \, .
\end{equation}
This is an implicit equation for detectable ELF signal power $\mathcal{E}^{95\%}_\mathrm{ELF}$.
The above equation can be represented in terms of the Marcum $Q$-function, which is a part of standard mathematical libraries,
\begin{equation}
    Q_{M}\prn{\sqrt{2  \mathcal{E}^{95\%}_\mathrm{ELF}} \,, \sqrt{2\mathcal{E}_{\mathrm{obs}}\,} }= 0.95 \,.
  \label{App:Eq:withMarcum}
\end{equation}

To find the sensitivity  to ELFs, we assume that the ELF signal is well below the noise floor. Then in Eq.~(\ref{App:Eq:withMarcum}), $\mathcal{E}_{\mathrm{obs}} \approx M$, see Eq.~(\ref{App:Eq:ExpVar}).   Inverting the resulting equation in the limit $M\gg 1$, we find
\begin{equation}
 \mathcal{E}^{95\%}_\mathrm{ELF} \approx 1.7\sqrt{M} \, .
 \label{App:Eq:E95}
\end{equation}
This result is consistent with the qualitative signal-to-noise ratio (SNR) arguments. SNR can be defined as
\[
 \mathrm{SNR} = \frac{  \mathcal{E}_\mathrm{ELF}}{ \sqrt{\text{Var}(\mathcal{E})}} = \frac{  
 \mathcal{E}_\mathrm{ELF}}{ \sqrt{M}} \, ,
\]
where we used Eq.~(\ref{App:Eq:ExpVar}) for the variance with only the noise contribution. Fixing the SNR value results in the same $\sqrt{M}$ scaling of the minimum detectable ELF power as in the more rigorous estimate  (\ref{App:Eq:E95}).

With these results, we can establish  sensitivity to coupling constants characterizing ELF portals. We parameterize
the ELF-induced  signals in the sensor as
\begin{equation}
 s(t)=\left\{
\begin{array}
[c]{cl}%
\gamma_1 \mathcal{C}_1 \phi(t)  \,, &  \text{linear} \label{App:Eq:Signals}\\
\gamma_2 \mathcal{C}_2 \phi(t)^2 \,, &  \text{quadratic}
\end{array}
\right. \,. 
\end{equation}
 Here $\gamma_1$ and $\gamma_2$ are coupling constants to be constrained and 
$\mathcal{C}_i$ are {\em known} constants determined by the particular sensor. 

Next we compute $\mathcal{E}_\mathrm{ELF}$,
the excess power statistic~(\ref{App:Eq:StatisticELF})  for the ELF signals~(\ref{App:Eq:Signals}). The  signal powers are  normalized to the noise PSD $\tilde{C}_p$. For a sensor exhibiting white noise of variance $\sigma^2$, the noise PSD is $\tilde{C}_p =N_\mathrm{w}  \sigma^2$ and
\begin{equation}
   \mathcal{E}_\mathrm{ELF}  =\frac{1}{N_\mathrm{w}  \sigma^2} \sum_{(p,w) \in \mathrm{ELF}} | \tilde{s}_{p,w}|^2  \, .
    \label{App:Eq:StatisticELF-white-noise}
\end{equation}
The sum over ELF contributions can be simply evaluated in the limit when the temporal window size $T_\mathrm{w}$ is  much smaller than  duration  of the  ELF burst $\tau$. Then  we can neglect the  time variation in the ELF envelope over the window. In the  window, the ELF frequencies span the frequency interval $ |d \omega/dt | T_\mathrm{w} = T_\mathrm{w}/(\tau  \tau_0)$, where the slope is given by Eq.~(\ref{App:Eq:Slope}).  Without loss of generality, we require that this spanned frequency interval is smaller than the DFT frequency resolution 
$\Delta_\omega = 2\pi /T_\mathrm{w}$.  We also require that adjacent windows map instantaneous ELF frequencies to distinct and adjacent DFT frequencies. Under these assumptions,  the total number $M$ of ELF-containing tiles  and the ``optimal'' window duration $T_\mathrm{w}$ are
\begin{eqnarray}
 M &\approx &\tau/T_\mathrm{w} \,, \label{App:Eq:NumTilesELF} \\
  T_\mathrm{w}& \approx& \sqrt{ 2 \pi \tau \tau_0} \, . \label{App:Eq:Tw}
\end{eqnarray}
With the negligible ELF frequency variation over the window, the field PSD  
\begin{equation}
|\tilde{\phi}_{w,p}|^2   \approx  \frac{1}{4} |\phi_\mathrm{env}(t_w)|^2 N_\mathrm{w}^2 \delta_{p,p_0} \, ,
\end{equation}
where $\phi_\mathrm{env}(t_w)$ is the value of the ELF burst envelope in the window and $p_0$ corresponds to the DFT frequency nearest to the ELF frequency in the window.  Summing over windows, we arrive at the minimal detectable ELF power
\begin{equation}
   \mathcal{E}_\mathrm{ELF,1}  \approx  \frac{\sqrt{\pi}}{4} \gamma_1^2  \mathcal{C}_1^2  \frac{1}{ \sigma^2} \frac{\tau}{\Delta_t}\phi_\mathrm{max}^2 
       \label{App:Eq:StatisticELFSimplifiedLin}
\end{equation}
 for the linear portal. To arrive at this result, we evaluated the sum   in the continuous limit,
 \begin{align*}
 \sum_{(p,w) \in \mathrm{ELF}} | \tilde{\phi}_{p,w}|^2 & \approx \frac{N_\mathrm{w}^2}{4} \sum_{w \in \mathrm{ELF}}  |\phi_\mathrm{env}(t_w)|^2 \\
 & \approx  \frac{N_\mathrm{w}^2}{4T_\mathrm{w}} \int_{-\infty}^{\infty}\phi_\mathrm{env}^2(t) dt
 \end{align*}
 and used the envelope for the Gaussian pulse.
 Similar evaluation for quadratic  portal leads to 
 \begin{equation}
   \mathcal{E}_\mathrm{ELF,2}  \approx  \sqrt{\frac{\pi}{2}} \gamma_2^2  \mathcal{C}_2^2  \frac{1}{16 \sigma^2} \frac{\tau}{\Delta_t}\phi_\mathrm{max}^4 \,.
       \label{App:Eq:StatisticELFSimplifiedQuad}
\end{equation}

Notice that for the quadratic coupling,  
\begin{align*}
&\cos^2\prn{ \omega_0 (t-t_s) - \frac{1}{2\tau \tau_0}(t-t_s)^2 }= \\
&\frac{1}{2}\sbrk{1+\cos\prn{ 2\omega_0 (t-t_s) - \frac{1}{\tau_0 \tau}(t-t_s)^2 }} \, ,
\end{align*}
i.e., the central frequency and the slope are doubled, while the field amplitude is effectively reduced by $\sqrt{2}$. We ignore the DC contribution in our present approach, although the DC contribution can serve as an additional signature for the quadratic interactions.

In  formulae~(\ref{App:Eq:StatisticELFSimplifiedLin},\ref{App:Eq:StatisticELFSimplifiedQuad}), the ratio $\tau/\Delta_t$ can be recognized as the total number of sampled points during the ELF pulse duration. These formulas together with the minimum detectable excess power~(\ref{App:Eq:E95}) yield the constraint on the coupling constant
\begin{equation}
\gamma_1^{95\%} \approx 2 \frac{\sigma}{\mathcal{C}_1 \phi_\mathrm{max} } \sqrt{\frac{\Delta_t}{\tau}}  
\prn{\frac{\tau}{\tau_0 }}^{1/8}
\end{equation}
 for the linear coupling and
\begin{equation}
\gamma_2^{95\%} \approx 4.7 \frac{\sigma}{\mathcal{C}_2 \phi_\mathrm{max}^2 } \sqrt{\frac{\Delta_t}{\tau}}  
\prn{\frac{\tau}{\tau_0 }}^{1/8}
\end{equation}
for  the quadratic coupling. Here we used the total number of ELF containing tiles~(\ref{App:Eq:NumTilesELF}) and the optimal window size~(\ref{App:Eq:Tw}). Since the ELF signal is coherent across a  sensor network,   the above constraints are improved by $\sqrt{N_s}$ for  a network of $N_s$ sensors (see more detailed discussion of statistical analysis for sensor networks in 
Refs.~\cite{Panelli:2019-MFT-GPSDM,Derevianko2016a,RomanoCornish2017}). 
Notice that the dependence on the ratio $\tau/\tau_0$ is weak and we drop this dependence.  Then with the maximum field amplitude~(\ref{App:Eq:PhiMax}),
\begin{align}
\gamma_1^{95\%} &\approx 6.5 \frac{\sigma }{\mathcal{C}_1 \sqrt{N_s} } R \omega_0 \sqrt{\frac{\Delta_t}{c \Delta E}}  \, , \label{App:Eq:LinGammaConstraint}\\
\gamma_2^{95\%} &\approx 52 \frac{\sigma }{\mathcal{C}_2  \sqrt{N_s} }  \frac{R^2 \omega_0^2}{c \Delta E} \sqrt{\Delta_t \tau}  \,. \label{App:Eq:QuadGammaConstraint}
\end{align}

These constraints  depend on the ELF central frequency $\omega_0$. The derivations in Appendix~\ref{App:Dispersion} are valid in the limit $\omega_0 \gg \Delta \omega= 1/\tau_0$. Then to avoid DFT
 aliasing, it is sufficient to  require that $\omega_0 \ll \pi/\Delta_t$, i.e., it is well below the Nyquist frequency. Or, explicitly,
\begin{equation}
1/\tau_0 \ll  \omega_0 \ll \pi/\Delta_t \, . \label{App:Eq:LimitsOnELFfrequency}
\end{equation}
While the upper limit is fixed by the sensor sampling rate, the initial ELF pulse duration $\tau_0$ depends on production mechanisms. For a general search with $\tau_0$ being a free parameter, the minimum detectable ELF frequency is on the order of the DFT (angular) frequency resolution, $2\pi/T_\mr{w}$. Considering that the typical rate of LIGO GW detections is a few events per year, we can take $T_\mr{w} \lesssim 10^6 \, \mr{s}$, leading to  
$(\omega_0)_\mr{min}\sim (2  \pi) \times 10^{-6}\, \mr{Hz}$.


\section{Atomic clocks and cavities}
\label{App:Sec:AtomicClocksCavities}
In Appendix~\ref{App:Sec:DataAnalysis}, we derived general constraints~(\ref{App:Eq:LinGammaConstraint},\ref{App:Eq:QuadGammaConstraint}) on linear and quadratic couplings to ELFs 
for a generic quantum sensor.  Here we specialize that discussion to atomic clocks and cavities. 

{\em Atomic clocks ---} 
Atomic clocks are quantum sensors which effectively compare frequency of an atomic transition with the resonance frequency of the local oscillator (LO).  The LO is typically a reference optical or microwave cavity.  The atoms (quantum oscillators) are interrogated with  laser or microwave pulses outcoupled from the cavities. The cavity frequency is tunable and a feedback  (servo) loop drives the LO frequency to be in resonance with the reference atomic transition. To tell time, 
the oscillations are counted at the source and converted to the time measurement by  multiplying the count with the fixed and known oscillation period of the quantum oscillator.   As cavity frequencies drift over time, locking LO frequencies to a stable atomic transition frequency is essential. Below we follow the simple model of atomic clock operation described in Ref.~\cite{Derevianko2016a} and  generalize it to the case of ELF detection.

In our preceding discussion, we assumed that the measurements were instantaneous; in practice, there is always a finite interrogation time $t_0$ for a single measurement. We assume that the next measurement is taken right after the previous 
one was completed. Then the DFT sampling time interval  $\Delta_{t} =t_{0}$. Typical interrogation time $t_0$ for modern atomic clocks is on the order of a second. In our simplified model of an atomic clock, we ignore the LO-quantum oscillator feedback loop. Feedback operations typically take a few measurement cycles and would attenuate rapid changes in the atomic/LO frequencies. Thus our analysis will hold in the limit  when the period of the ELF oscillations
is larger than the interrogation time, i.e.,  $1/\omega_0 \gg t_0$.  This requirement 
is consistent with the DFT aliasing limit [upper limit  in Eq.~(\ref{App:Eq:LimitsOnELFfrequency})].

Modern atomic clocks measure the quantum phase $\Phi$ of an atomic oscillator with respect to the local oscillator. 
The ELF-induced accumulated phase difference is
\begin{align}
 \Phi_j^\mr{ELF}&= 2\pi \int_{t_{j-1}}^{t_j} [ \nu_\mr{atom}^\mr{ELF}(t') - \nu_\mr{LO}^\mr{ELF} (t')] dt'  \nonumber\\
  &\approx 2\pi [\nu_\mr{atom}^\mr{ELF}(t_j) - \nu_\mr{LO}^\mr{ELF} (t_j)] t_0 \,,
\end{align}
 since the observable ELF oscillations are slow over the interrogation time, cf. Eq.~(\ref{App:Eq:LimitsOnELFfrequency}).  The resulting frequency difference  is  typically recorded as an error signal by the servo-loop. Thereby, we consider a time series of fractional frequency excursions  
\begin{equation}
s_j \equiv \frac{ \nu_\mr{atom}(t_j) - \nu_\mr{LO}(t_j) }{\nu_\mr{clock}} 
\label{App:Eq:ClockErrorSignal}
\end{equation} 
taken at $t_j=j t_0; j= 1,2,,\ldots N_\mathrm{tot}$, with $\nu_\mr{clock}$  being the unperturbed clock frequency. 

Atomic and  cavity frequencies can be affected by  varying fundamental constants,  such as the fine structure constant $\alpha= e^2/\hbar c$ and/or fermion masses $m_f$. We consider a model where an ELF field drives such variations. Formally, 
these result from the following phenomenological Lagrangians  (portals) that couple standard model (SM)  fields and ELFs
\begin{eqnarray}
\mathcal{L}_{\rm{int}}^{(1)} = \prn{-\sum_f\Gamma_f^{(1)}  m_f c^2\bar{\psi}_f\psi_f + \frac{\Gamma_\alpha^{(1)}}{4} F_{\mu\nu}^2}\sqrt{\hbar c}\,\phi\,, \label{App:Eq:L1clocks} \\
\mathcal{L}_{\rm{int}}^{(2)} = \prn{-\sum_f\Gamma_f^{(2)} m_f c^2\bar{\psi}_f\psi_f + \frac{\Gamma_\alpha^{(2)}}{4} F_{\mu\nu}^2} \hbar c\,\phi^2\,. \label{App:Eq:L2clocks}
\end{eqnarray}
$\mathcal{L}_{\rm{int}}^{(1)}$ is linear in the exotic field $\phi$, while $\mathcal{L}_{\rm{int}}^{(2)}$ is quadratic.
Here we used the Lorentz-Heaviside system of electromagnetic units that is common for particle physics literature.
The structure of these portals is such that various parts of the  SM Lagrangian are multiplied by exotic fields,
with $\Gamma$'s being the associated coupling constants (to be determined or constrained).
In the above interactions, $f$  runs over all the SM fermions (fields $\psi_f$ and masses $m_f$),  and $F_{\mu\nu}$ is the Faraday tensor; one may include gluon, Higgs, or weak interaction contributions if desired.
We refer the interested reader to the discussion of  technical naturalness of such Lagrangians in Ref.~\cite{derevianko2014hunting}. 
 In these expressions, the combination $\sqrt{\hbar c}\,\phi$ is measured in units of energy, $[E]$. Then  
 $\Gamma_X^{(1)}$ are measured in $[E]^{-1}$ and $\Gamma_X^{(2)}$ --- in $[E]^{-2}$. 

 The portals~(\ref{App:Eq:L1clocks}) and (\ref{App:Eq:L2clocks}) lead to the effective redefinition of fermion masses and the fine-structure constants:
\begin{gather}
m_f(\mb{r},t) = m_f \times\sbrk{1+ \Gamma_f^{(n)} \prn{\sqrt{\hbar c}\, \phi(\mb{r},t)  }^n }\,,\nonumber
\\ 
\alpha(\mb{r},t) \approx  \alpha \times\sbrk{1+ \Gamma_\alpha^{(n)} \prn{\sqrt{\hbar c} \, \phi(\mb{r},t)  }^n }\,,
\end{gather}
for the linear ($n=1$) and quadratic  ($n=2$) portals, where $m_f$ and $\alpha$ are the nominal (unperturbed) values.

Atomic frequencies are primarily affected by the induced variation of the Rydberg constant, $\mathcal{R}_\infty = m_e c^2 \alpha^2$. Optical clocks can exhibit additional $\alpha$ dependence due to relativistic effects. Microwave clocks operate on hyperfine transitions and are additionally affected by the variation in the quark masses, $m_q$ and the strong coupling constant.   The reference cavity is also a subject to the ELF influence. For example, the variation in the Bohr radius $a_{0} =  \alpha^{-1} \hbar/(m_{e}c)$ affects cavity length $L \propto a_{0}$ and thus the cavity resonance frequencies~\cite{StaFla2015,Wcislo2016,Roberts2017-GPS-DM,Derevianko2016a}.  Conventionally, one introduces
 coefficients  $\kappa_X=\partial \ln \nu/\partial \ln X$ quantifying sensitivity of a resonance frequency $\nu$   to the variation in the fundamental constant $X$.  Then 
 \begin{align*}
 \kappa_{m_e}^\mr{atom}  &\approx   1 \,,\\ 
 \kappa_{\alpha}^\mr{atom} & \approx  2 \,,\\
 \kappa_{m_e}^\mr{cavity}  &\approx -1  \,,\\ 
 \kappa_{\alpha}^\mr{cavity} & \approx -1  \,.
 \end{align*}
 It is worth noting that there are exceptional cases of enhanced sensitivity to variation of fundamental  constants,
 for example, in actively pursued, but yet not demonstrated,  $^{229}$Th nuclear clock~\cite{CamRadKuz12} ($\kappa_\alpha \approx 10^4 $, Ref.~\cite{Litvinova2009}), and clocks based on highly-charged ions~\cite{DerDzuFla12} ($\kappa_\alpha \lesssim 10^2$, Ref.~\cite{DzuFla2015-VarHCI-Review}).
 The above arguments presuppose instantaneous adjustment of the resonance/transition frequencies to  the variation of fundamental constants, see Ref.~\cite{Derevianko2016a} for further discussion. 
 
 The sought ELF signal~(\ref{App:Eq:ClockErrorSignal})  is expressed in terms of the differential sensitivity coefficient $K_X= \kappa_X^\mr{atom} -\kappa_X^\mr{LO}$,
 \begin{equation}
s_j =  \Gamma_\mr{eff}^{(n)} \prn{\sqrt{\hbar c}\, \phi(t_j)  }^n \, ,
\label{App:Eq:ClockSignalELFs}
\end{equation}
where $n=1$ or $2$ for the linear and quadratic portals respectively.
Here we introduced the effective coupling constants
\begin{equation}
\Gamma_\mr{eff}^{(n)} \equiv \sum_X{ K_X \Gamma_X^{(n)} } , \label{App:Eq:EffectiveGamma}
\end{equation}
with the sum over all relevant fundamental constants. Comparing Eq.~(\ref{App:Eq:ClockSignalELFs}) with our  generic ELF signal template~(\ref{App:Eq:Signals}) leads to the identification $\gamma_n =\Gamma_\mr{eff}^{(n)} $ and $\mathcal{C}_n = \prn{\hbar c}^{n/2}$. To apply the derived constraints~ (\ref{App:Eq:LinGammaConstraint},\ref{App:Eq:QuadGammaConstraint}), 
 we also need to make  an assumption about the nature of the measurement noise, which for atomic clocks is characterized by the  Allan deviation $\sigma_{y}(\tau_\mr{meas})$, where $\tau_\mr{meas}$ is the measurement time. If the Allan deviation scales as $\sigma_{y}(\tau_\mr{meas}) \propto 1/\sqrt{\tau_\mr{meas}}$, the measurement noise is dominated by the white  frequency noise. Then in  constraints~(\ref{App:Eq:LinGammaConstraint},\ref{App:Eq:QuadGammaConstraint}) $\sigma = \sigma_y(t_0) = \sigma_y(\Delta_t)$ and  we immediately arrive at constraints on the effective coupling constants (at  the 95\% C.L.)
 \begin{align}
\Gamma_\mr{eff}^{(1)} & \lesssim 6.5 
\frac{\sigma_y(\Delta_t) }{\sqrt{N_s} } \prn{  \frac{\omega_0}{c} R} \prn{\frac{\Delta_t}{\hbar \Delta E}}^{1/2}  \, , \label{App:Eq:Clocks:LinGammaConstraint}\\
\Gamma_\mr{eff}^{(2)} &\lesssim 52 \frac{\sigma_y(\Delta_t) }{ \sqrt{N_s} }  \prn{  \frac{\omega_0}{c} R}^2  \frac{1}{ \Delta E} \prn{\frac{\Delta_t \tau}{\hbar^2} }^{1/2}  \,. \label{App:Eq:Clocks:QuadGammaConstraint}
\end{align}

{\em Optical cavities ---}
Atomic clocks have a relative low $\sim \mr{Hz}$ sampling rate. Terrestrial networks of such clocks would not be able to track propagation of the ultra-relativistic ELF pulse through the network as discussed in the main text. One of the  possibilities is to employ a network of optical cavities providing a much higher, $\gtrsim 10\,  \mr{kHz}$, sampling rate. Each node would contain two distinct cavities: one with a rigid spacer and the other with suspended mirrors (without the spacer, similar to LIGO cavities). The resonance frequency of the cavity with a rigid spacer is affected by the variation of fundamental constants, while that of the cavity without the spacer is not. The experiment would involve comparison  of these resonance frequencies. This scheme was proposed in the context of the search for ultralight dark matter~\cite{Cavity.DM.2018}, and can be adopted for the ELF searches.
The   constraints~(\ref{App:Eq:LinGammaConstraint}) and (\ref{App:Eq:QuadGammaConstraint}) immediately apply with $\Gamma_\mr{eff}^{(n)}$, Eq.~(\ref{App:Eq:EffectiveGamma}), involving sensitivity coefficient of the rigid spacer cavity:
$K_X = \kappa^\mr{cavity}_X$. Another related high sampling rate possibility is the three-arm Mach-Zender interferometer~\cite{Savalle2019-DAMNED}, where the delays of laser pulse are compared while traveling through an optical cavity and an optical fiber.

{\em Linear couplings---}
 Here we focus on the linear coupling and assume for simplicity that one of the coupling dominates,  
 e.g., $\Gamma_\mr{eff}^{(1)}\approx K_\alpha \Gamma^{(1)}_{\alpha}$. This assumption is hardly necessary but it clarifies the role of the sensitivity coefficients $K_X$.
 We recast the constraint~(\ref{App:Eq:Clocks:LinGammaConstraint})  in terms of  moduli~\cite{dilaton-limits}  $d_X \equiv (E_\mr{Pl}/\sqrt{4\pi})\Gamma^{(1)}_X$, with  $E_\mr{Pl}=\sqrt{\hbar c^5/G}$ being the Planck energy. 
  \begin{equation}
d_X \lesssim 1.8 \,\frac{E_\mr{Pl}}{K_X } 
\frac{\sigma_y(\Delta_t) }{\sqrt{N_s} } \prn{  \frac{\omega_0}{c} R} \prn{\frac{\Delta_t}{\hbar \Delta E}}^{1/2}  \, .\label{App:Eq:ModuliConstraint}
\end{equation}
or, in practical units,
\begin{widetext}
 \begin{equation}
d_X \lesssim 54 \,\frac{1}{K_X \sqrt{N_s}} 
\prn{\frac{\sigma_y(\Delta_t) }{ 10^{-16}} }
\prn{  \frac{\omega_0}{2 \pi \,\mr{Hz}} } 
\prn{ \frac{R}{\mr{Mpc}} }  
\prn{\frac{\Delta_t}{\mr{s}} }^{1/2}   \prn{ \frac{ \Delta E }{\Msun}}^{-1/2}  \, .\label{App:Eq:ModuliConstraintPractical}
\end{equation}
\end{widetext}
Here, as the reference value for the Allan deviation, we took $\sigma_y(1\,\mr{s}) \approx 10^{-16}$ characteristic of modern optical lattice clocks~\cite{LudBoyYe15-OpticalClocks-review}.

We  focus on the electron mass modulus $d_{m_e}$ and the electromagnetic gauge modulus $d_e$ ($X=\alpha$ in this case). The most stringent limits on these moduli come from equivalence principle violation tests (see Fig.~1 of Ref.~\cite{dilaton-limits}). For the parameter space relevant to clocks and cavities, the excluded regions are $d_e \gtrsim 10^{-3}$ and 
$d_{m_e} \gtrsim 10^{-2}$.

{\em Quadratic couplings ---}
For consistency with prior literature, we rewrite the constraint~(\ref{App:Eq:Clocks:QuadGammaConstraint}) in terms of the  energy scale $\Lambda_X = 1/\sqrt{|\Gamma_X^{(2)}|}$ ,
 \begin{equation*}
\Lambda_X \gtrsim 0.14 \, \sqrt{|K_X|}  
\prn{ \frac{\sqrt{N_s} }{\sigma_y(\Delta_t) }}^{1/2}
 \prn{  \frac{c}{R\omega_0}} 
 {\Delta E}^{1/2}  \prn{\frac{\hbar^2} {\Delta_t \tau}}^{1/4} 
  \, .\label{App:Eq:LambdaConstraint}
\end{equation*}
Here we assumed that the variation in a fundamental constant $X$ dominates (say, $\Gamma_\mr{eff}^{(2)}
 \approx K_{m_e} \Gamma_{m_e}^{(2)}$). In practical units,
 \begin{widetext}
\begin{equation}
\frac{\Lambda_X}{\mr{TeV}} 
\gtrsim  1.8 \times 10^5   \, |K_X|^{1/2}  N_s^{1/4} \times  
 \prn{ \frac{\sigma_y(\Delta_t) }{ 10^{-16}}}^{-1/2}
\prn{  \frac{R}{\mr{Mpc}} \times \frac{\omega_0}{2 \pi \,\mr{Hz}} } ^{-1} 
\prn{ \frac{ \Delta E }{\Msun}}^{1/2} 
\prn{\frac{\Delta_t}{1\, \mr{s} } \times \frac{\tau}{10^2\, \mr{s}}}^{-1/4}
  \, .\label{App:Eq:LambdaConstraintPractical} 
\end{equation}
\end{widetext}
The most stringent constraints on the energy scales 
\begin{equation}
    \Lambda_{m_e,\alpha} \gtrsim 3 \, \mathrm{TeV}  \text{ and } \Lambda_{m_p} \gtrsim 10 \, \mathrm{TeV}
    \label{Eq:quad-limits}
\end{equation}
come from the bounds on the thermal emission rate from the cores of supernovae~\cite{Olive:2007aj}. These authors analyzed emissivity of $\phi$ quanta due to pair annihilation of photons and other processes such as the bremsstrahlung-like emission. 
They also considered tests of the gravitational force which result in similar constraints; compared to linear Lagrangians these are mild, because the quadratic Lagrangians lead to the interaction potentials that scale as an inverse {\em cube} of the distance as only the exchange of pairs of $\phi$'s are allowed (for linear Lagrangians, the $\phi$-mediated interaction potentials scale as the inverse distance).

From the numerical pre-factor in Eq.~(\ref{App:Eq:LambdaConstraintPractical}), it is clear that a generic ELF search would probe energy scales well beyond the existing astrophysical and gravity test bounds. This is  further illustrated in Fig. 3 of  the main text.

 \section{Magnetometers}
 Atomic magnetometer measure the response of atomic magnetic moments to magnetic fields.
We consider interaction Lagrangians~\cite{pospelov2013detecting} that are linear, $\mathcal{L}^{(1)}$, and quadratic, $\mathcal{L}^{(2)}$, in the spin-0 ELF fields $\phi$,
 \begin{align}
\mathcal{L}^{(1)}_\mr{mag} &= f_l^{-1} J^\mu \partial_\mu \phi \, , \\
 \mathcal{L}^{(2)}_\mr{mag} &= f_q^{-2} J^\mu \partial_\mu \phi^2\, .
\end{align}
In these expressions, $J^\mu = \bar{\psi} \gamma^\mu \gamma_5 \psi$ is  the axial-vector current for SM fermions and  $f_l$, $f_q$ are the characteristic energy scales associated with the linear and quadratic spin portals, respectively. The relevant contribution to the Dirac Hamiltonian can be computed as 
\begin{equation}
H_\mr{int}\psi=-\gamma_{0}\left(  \frac{\partial\mathcal{L}_\mr{int}}%
{\partial\bar{\psi}}-\partial_{\mu}\left(  \frac{\partial\mathcal{L}_\mr{int}%
}{\partial\left(  \partial_{\mu}\bar{\psi}\right)  }\right)  \right)  \,,
\end{equation}
leading to
\begin{align}
H^{(1)}_\mr{mag} &= -
\frac{1}{f_l} \prn{  \gamma_{5} \frac{\partial}{c\partial t}\phi
+  \mb{\Sigma} \cdot \bm{\nabla} \phi } \, , \label{Eq:Lin-Mag-Hamiltonian} \\
H^{(2)}_\mr{mag} &= -\frac{1}{f_q^2} \prn{  \gamma_{5} \frac{\partial}{c\partial t}\phi^2
+  \mb{\Sigma} \cdot \bm{\nabla} \phi^2 } \,. \label{Eq:Quad-Mag-Hamiltonian}
\end{align}
Here we used identities $\gamma_0 \gamma_0  = 1$ and $\gamma_0 \gamma^i \gamma_5  = \Sigma^i$ with  the spin matrix
\begin{equation}
\mathbf{ \Sigma} =\left(
\begin{array}[c]{cc}
\bm{\sigma} & 0 \\
   0                     & \bm{\sigma} 
\end{array} \right) \,.
\end{equation}

Atomic magnetometers, such as those employed in GNOME \cite{afach2018characterization}, are sensitive to spin-dependent energy shifts. Computing the expectation value of these Hamiltonians, we arrive at the effective spin-dependent interactions:
\begin{align}
H^{(1)}_\mr{mag} & \approx -
\frac{2 (\hbar c)^{3/2} }{f_l} \mb{S} \cdot \bm{\nabla} \phi  \, , \label{Eq:magnetometer-nonrel-Hamiltonians-lin} \\
H^{(2)}_\mr{mag} & \approx -
\frac{2 (\hbar c)^{2}}{f_q^2} \mb{S} \cdot \bm{\nabla} \phi^2  \, , \label{Eq:magnetometer-nonrel-Hamiltonians-quad}
\end{align}
equivalent to the non-relativistic Hamiltonians often seen in the literature (see, e.g., Ref.~\cite{safronova2018search}). The terms containing time derivatives of the $\phi$ field are neglected in the non-relativistic limit for atomic electrons or nucleons as the $\gamma_5$  matrix mixes large and small components of the Dirac bi-spinors.  $\bm{S}$ is the atomic or nuclear spin. 

The ELF Hamiltonians described by Eqs.~\eqref{Eq:magnetometer-nonrel-Hamiltonians-lin} and \eqref{Eq:magnetometer-nonrel-Hamiltonians-quad} can be related to the general forms of the ELF interactions given in Eq.~\eqref{App:Eq:Signals} through the following identifications:
\begin{align}
    \gamma_1 &= -\frac{1}{f_l} \, , \\
    \mathcal{C}_1 &\approx 2\hbar^{3/2}c^{1/2}\omega_0 \, , \\
    \gamma_2 &= -\frac{1}{f_q^2} \, , \\
    \mathcal{C}_2 &\approx 4\hbar^{2}c\omega_0 \, ,
\end{align}
where we have kept only the leading terms when taking the gradients of $\phi$ and $\phi^2$. Note that one must also take into account the atomic and nuclear structure \cite{kimball2015nuclear} as well as geometrical considerations \cite{afach2018characterization} to interpret magnetometer data in terms of couplings to ELFs, but for the rough estimates presented in this work we ignore these details. With these identifications, from Eqs.~(\ref{App:Eq:LinGammaConstraint},\ref{App:Eq:QuadGammaConstraint}) we arrive at the constraints on the effective coupling constants (at the 95\% C.L.):
\begin{align}
    f_l &\gtrsim \frac{\hbar^{3/2}c}{3} \frac{\sqrt{N_s}}{\sigma_m(\Delta_t) \sqrt{\Delta_t}} \frac{\sqrt{\Delta E}}{R} \, , \\
    f_q^2 &\gtrsim \frac{\hbar^2c^2}{13} \frac{\sqrt{N_s}}{\sigma_m(\Delta_t) \sqrt{\Delta_t \tau}} \frac{\Delta E}{R^2\omega_0} \, .
\end{align}
Here $\sigma_m(\Delta_t)$ is the magnetometer energy resolution.
A typical GNOME magnetometer has a bandwidth of $\approx 100\,{\rm Hz}$ and, integrating over a time $\Delta_t$, can measure the magnetic field with precision given by $\delta B~\approx 100\,
\mathrm{fT}  \sqrt{\mathrm{s}}/\sqrt{\Delta_t}$ \cite{afach2018characterization}. Thus
\begin{align}
\label{Eq:energy-resolution-GNOME}
    \sigma_m(\Delta_t) \approx g \mu_B \delta B \approx \frac{10^{-18}}{\sqrt{\Delta_t}}~{\rm{eV}\sqrt{s}}~,
\end{align}
where $g$ is the gyromagnetic ratio (which depends on the atomic species used in the magnetometer) and $\mu_B$ is the Bohr magneton. The prior astrophysical limits on energy scales are $f_l \approx 2 \times 10^8~{\rm GeV}$ \cite{Chang2018} and $f_q \approx 10^4~{\rm GeV}$ \cite{pospelov2013detecting}.

\section{Astrophysical reach of existing/planned sensor networks}

\begin{table*}[ht]{Clock Network Sensitivity Estimates (Linear Portal)}
\begin{tabular}{lccccc}
\hline\hline
\rule{0ex}{3.0ex} Clock Network & ~Allan deviation~ & ~Astrophysical reach~ & ~Volume probed~ & ~Event rate~ \\
\rule{0ex}{3.0ex}  & $\sigma_y(1\rm{s})$ & [ly] & [Gpc$^3$] & [1/yr] \\
\hline
\rule{0ex}{3.0ex} GPS                                  & $10^{-13}$    & $10^3$      & $10^{-19}$   & -   \\
\rule{0ex}{3.0ex} Optical lattice clocks      & $10^{-16}$    & $10^5$      & $10^{-11}$   & $10^{-6}$   \\
\rule{0ex}{3.0ex} Th nuclear clocks ($\star\star$)     & $10^{-14}$    & $10^{8}$      & $10^{-3}$    & $1$   \\
\hline
\end{tabular}
\caption{Estimated sensitivity to ELFs for {\em linear} couplings, astrophysical reach, volume probed, and ELF event rates for the linear couplings to atomic clocks. Estimates are carried out for existing and theoretically possible ($\star\star$) atomic clock sensor networks. Event rates assume an ELF energy release of $\Delta E \approx M_\odot c^2$ and a generic binary merger rate density of $10^3~{\rm Gpc^{-3}yr^{-1}}$. Here we use Allan deviations for a 1~Hz sampling rate and electromagnetic gauge modulus $d_e=10^{-3}$. For reference, the observable universe has a volume of $10^4$ Gpc$^3$.}
\label{tab:clock-event-rates}
\end{table*}

{\em Atomic clocks ---}
GPS is a network that is comprised of nominally 32 satellites in medium-Earth orbit (altitude $\sim 20,000$ km) and functions by using atomic clock transitions (based on either Rb or Cs atoms) to drive microwave signals which are broadcast to Earth \cite{roberts2017search,roberts2018search}. A network of specialized Earth-based GPS receivers measures the carrier phase of these microwave signals which is then used in the processing required to produce the GPS clock time-series data.  Due to the network's advantageous spatial extent, the clocks on-board the GPS satellite constellation are used to comprise the network of precision measurement sensors, but the network can also include the $\sim 40$ high-quality Earth-based receiver stations, several of which use highly-stable H-maser clocks, along with Rb, Cs, and quartz oscillators \cite{roberts2018search}. Due to their better noise characteristics,  recent satellites in the constellation predominantly use Rb based clocks. As of August 2018, there were 30 Rb satellites and only one Cs satellite in operation. The GPS satellites are grouped into several version generations, called blocks: II, IIA, IIR, and IIF, with Block III currently under development. Newer generation satellites have improved noise characteristics of the satellite clock network~\cite{roberts2018search}. The network can be extended to incorporate  clocks from other Global Navigation Satellite Systems, such as the European Galileo, Russian GLONASS, and Chinese BeiDou, and networks of laboratory clocks.
 
Normally the GPS network data, as provided by the Jet Propulsion Laboratory (JPL), has a $ \Delta_t=30\, \mr{s}$ sampling time interval, but many Earth-based receivers probe the satellite signals at a higher rate. Search for ELFs calls for higher sampling rate in the generated clock data and recently the GPS.DM collaboration produced 1 Hz  rate satellite clock data.
This is the reason that we used $\Delta_t= 1\, \mr{s}$ in the main text and below. Such sampling time allows us to probe ELF frequencies up to the Nyquist frequency, $0.5 \, \mr{Hz}$. Notice that  $\Delta_t= 1\, \mr{s}$ is still not fast enough to resolve a light-speed propagation event across the constellation even with the large $\sim 50,000 \, \mr{km}$  spatial extent of the satellite network, as a light-speed pulse would only be within the network for $\sim~0.2 \,\textrm{s}$. Thus we treat the GPS network as one collective sensor for ELF search.

The sensitivity to linear coupling constants is given by Eq.~(\ref{App:Eq:ModuliConstraintPractical}). Alternatively, one could use a fixed value for $d_e$ based on equivalence principle violation constraints $d_e<10^{-3}$ \cite{dilaton-limits}, and solve for the maximum astrophysical range $R$. 
If we pick optimal values for the parameters in Eq.~(\ref{App:Eq:ModuliConstraintPractical}), this can function as a maximum sensitivity for the clock networks for the linear coupling case. For Rb GPS clocks, the sensitivity coefficient is $K_\alpha=2$, and they have a typical Allan deviation $\sigma_{y}(1\,\mr{s}) \approx 10^{-13}$. This leads to an astrophysical range of $\approx 10^4 \,\mr{ly}$ for a detector network of $N_s \sim 100$ clocks, which is achievable with the incorporation of other satellite positioning networks. Optical clock  networks have a much better Allan deviation~\cite{hinkleyYbinstability, Jiang2011MakingStabilization} and can reach farther than $\approx 10^5$~ly, encompassing entire Milky Way. Potential future $^{229}\mr{Th}$  nuclear clocks have a much higher projected sensitivity coefficient $K_\alpha \approx 10^4$~\cite{thorium-coupling}, and an Allan deviation $\sigma_{y}(1\mr{s}) \approx 10^{-14}$~\cite{kazakov2012performance}. Nuclear clocks will allow for a maximum range of $\approx 10^8$~ly, which is enough range to search for ELFs originating from sources as distant as the neutron star merger event GW170817. These estimates are reflected in Table~\ref{tab:clock-event-rates}.

The sensitivity to {\em quadratic} couplings is given by Eq.~(\ref{App:Eq:LambdaConstraintPractical}). The constraints on quadratic couplings are  more relaxed than for the linear case (see Sec.~\ref{App:Sec:AtomicClocksCavities} ), allowing for probing much larger unconstrained parameter space. Using $\Lambda_\alpha\gtrsim 3\,\mathrm{TeV}$ from Eq.~(\ref{Eq:quad-limits}) and using the same parameters as in the above discussion of the linear portal, we can compare the current limits with the best case sensitivity. We fix  $\tau\sim100\,\mathrm{s}$. For GPS Rb clocks, ELFs can be probed up to energy scales of $\Lambda_\alpha \sim 10^4 \,\mathrm{TeV}$. Optical lattice clock  networks can probe energy scales up to $\Lambda_\alpha \sim 10^5 \,\mathrm{TeV}$ and nuclear clocks up to $\Lambda_\alpha \sim 10^7 \,\mathrm{TeV}$. All of these clocks have a potential discovery reach encompassing the entire observable Universe.

\begin{table*}[ht]{Magnetometer Sensitivity Estimates}
\begin{tabular}{lccccc}
\hline\hline
\rule{0ex}{3.0ex} Magnetometer Network & ~Allan deviation~ & ~Astrophysical reach~ & ~Volume probed~ & ~Event rate~ \\
\rule{0ex}{3.0ex}  & ~~~$\sigma_m(1\rm{s})$ [eV]~~~ & ~[ly]~ & ~[Gpc$^3$]~ & ~[1/yr]~ \\
\hline
\rule{0ex}{3.0ex} GNOME                             & $10^{-18}$    & $10^3$ $(10^6)$        & $ 10^{-19}$ $(10^{-10})$           & - $(10^{-5})$ \\
\rule{0ex}{3.0ex} Advanced GNOME ($\star$)          & $10^{-20}$    & $10^5$ $(10^8)$        & $ 10^{-13}$ $(10^{-4})$                 & $10^{-6}$ $(0.1)$           \\
\rule{0ex}{3.0ex} Ferromagnetic gyro ($\star\star$)  & $10^{-25}$    & $10^{10}$ $(10^{11})$         & $10^{2}$ $(10^4)$                 & $10^5$ $(10^7)$    \\
\hline\\
\end{tabular}
\caption{Estimated sensitivity to ELFs, astrophysical reach, volume probed, and ELF event rates for the linear (quadratic) couplings to magnetometers. Estimates are carried out for existing, planned ($\star$), and theoretically possible ($\star\star$) magnetometer sensor networks. Event rates assume an ELF energy release of $\Delta E \approx M_\odot c^2$ and a generic binary merger rate density of $10^3~{\rm Gpc^{-3}yr^{-1}}$. Here we assumes and ELF-spin linear coupling constant $f_l \approx 2 \times 10^8~{\rm GeV}$ and quadratic coupling constant $f_q \approx 10^4~{\rm GeV}$. For reference, the observable universe has a volume of $10^4$ Gpc$^3$, saturated in the case of theoretically possible ($\star\star$) sensor networks for the quadratic interaction. In the case of the quadratic coupling, a Fourier-limited signal with $\omega_0 \sim 10^{-4}~{\rm s^{-1}}$ is assumed.}
\label{tab:mag-event-rates}
\end{table*}

{\em Magnetometers ---} The astrophysical reach for a network of atomic magnetometers can be estimated based on the sensitivity of the magnetometers to spin-dependent energy shifts. The GNOME is just such a network of shielded optical atomic magnetometers specifically targeting transient events associated with beyond standard model physics \cite{pospelov2013detecting,afach2018characterization,pustelny2013global,kimball2018searching,GNOMEwebsite}. Presently GNOME consists of $N_s = 12$ dedicated optical atomic magnetometers \cite{budker2013optical} located at stations throughout the world (six sensors in North America, three in Europe, and three in Asia), with a number of new stations under construction in Israel, India, Australia, and Germany \cite{GNOMEwebsite}. Each magnetometer is located within a multi-layer magnetic shield to reduce the influence of magnetic noise and perturbations while retaining sensitivity to exotic spin-dependent interactions associated with beyond standard model physics \cite{kimball2016magnetic}, such as an ELF. The astrophysical reach of a GNOME-based search for ELFs using the spin portals can be estimated based on Eqs.~\eqref{Eq:LinSpinConstraint}, \eqref{Eq:QuadSpinConstraint}, and \eqref{Eq:energy-resolution-GNOME}, and is presented in Table~\ref{tab:mag-event-rates}.

In the near term, several stations around the world are upgrading their GNOME sensors to employ a dense polarized noble gas and a comagnetometer configuration, an experimental technique to search for beyond standard model physics pioneered by Romalis and coworkers \cite{kornack2005nuclear,vasilakis2009limits,brown2010new}. The new global network of noble gas comagnetometers will form an Advanced GNOME with an anticipated energy resolution a hundred times better than the existing GNOME, significantly increasing the astrophysical reach (Table~\ref{tab:mag-event-rates}). Finally, we note that there is ongoing long-term development of magnetometers based on levitated precessing ferromagnetic needle gyroscopes \cite{kimball2016precessing,band2018dynamics,wang2019dynamics,gieseler2019single,vinante2019ultrahigh}, a technology that, in principle, could improve energy resolution by a factor of $\sim 10^7$ compared to GNOME. These potential sensitivity improvements are noted in Table~\ref{tab:mag-event-rates}.

For numerical estimates of potential astrophysical range explored, we assume (1) an ELF energy release of $\Delta E \approx M_\odot c^2$ for BBH mergers and $\Delta E \approx 0.1 M_\odot c^2$ for BNS mergers (see reasoning in the main text), and (2) the maximum spin-dependent couplings consistent with existing astrophysical limits: $f_l \approx 2 \times 10^8~{\rm GeV}$ \cite{Chang2018} and $f_q \approx 10^4~{\rm GeV}$ \cite{pospelov2013detecting}. For BBH mergers, Advanced GNOME will have an astrophysical reach for linear couplings of $\approx 10^5$ light years, covering the entire Milky Way, and for quadratic couplings the astrophysical reach could be as large as $\approx 10^8$ light years. For neutron-star mergers, the respective astrophysical reach is reduced by a factor of $\approx 3$ due to the smaller $\Delta E$.  The present GNOME has a hundred times smaller astrophysical reach as compared to Advanced GNOME.


\section{ ELF event rates}
\label{App:Sec:event-rates}
The starting point for estimating the ELF burst rate is to determine the number of relevant astrophysical events in a given cosmic volume. In our case, we include BBH mergers, BNS mergers , and mergers of black hole with a neutron star (BH+NS), although ELF bursts may also come from other sources. Recent studies \cite{abbott2016rate,abbott2017gw170817,ali2017merger,mapelli2018cosmic,belczynski2018binary,chruslinska2017double,eldridge2018consistent} based on observed GW events estimate the binary merger rates may be as large as $\gamma({\rm BBH}) \sim 200~{\rm Gpc^{-3}yr^{-1}}$, $\gamma({\rm BH+NS}) \sim 3000~{\rm Gpc^{-3}yr^{-1}}$, and $\gamma({\rm BNS}) \sim 5000~{\rm Gpc^{-3}yr^{-1}}$. We conclude that it is reasonable to assume a generic binary merger rate of $\gamma \sim 10^3~{\rm Gpc^{-3}yr^{-1}}$.

A cosmic volume of $1~{\rm Gpc^3}$ contains roughly $10^9$ galaxies, so based on the above estimate for the merger rate $\gamma$, the rate of binary mergers in the Milky Way is $\sim 10^{-6}~{\rm yr^{-1}}$. This, for example, yields the expected event rate of Advanced GNOME for linear couplings, to have an astrophysical reach covering the entire Milky Way (Table~\ref{tab:mag-event-rates}). The same argument also yields the expected rate for a multi-network configuration of the GPS and Galileo satellite constellations (Table~\ref{tab:clock-event-rates}). Increasing the sensitivity of magnetometers and clocks has a dramatic impact on event rates: once a significant number of galaxies are within the astrophysical reach of the network, the cosmic volume probed becomes proportional to the cube of the sensor sensitivity.

Binary merger event rates within the Milky Way are $\ll 1/{\rm yr}$, and so it is exceedingly unlikely that GNOME or GPS will be able to detect an ELF burst coupled through the linear interaction correlated with a GW event in their current state of operation. The situation is more optimistic for ELFs coupled via the quadratic interaction as discussed in the main text. Future technologies \cite{kimball2016precessing,band2018dynamics,kazakov2012performance,von2016direct} offer the possibility of quantum sensor networks with much greater sensitivity and greater astrophysical reach.

\bibliographystyle{naturemag_noURL}
\bibliography{ELF,library-apd_AW}

\end{document}